# Estimands and Their Implications for Evidence Synthesis for Oncology: A Simulation Study of Treatment Switching in Meta-Analysis

**Authors**

Rebecca K Metcalfe,[1,2] Antonio Remiro-Azócar,[3] Quang Vuong,[1] Anders Gorst-Rasmussen,[4] Oliver Keene,[5] Shomoita Alam,[1] Jay JH Park[1,6]

*Affiliations*

1. Core Clinical Sciences Inc. Vancouver, BC, Canada
2. Centre for Advancing Health Outcomes, University of British Columbia, Vancouver, BC, Canada
3. Methods and Outreach, Novo Nordisk Pharma, Madrid, Spain
4. Novo Nordisk A/S, Søborg, Denmark
5. KeeneONStatistics, Maidenhead, UK
6. Department of Health Research Methodology, Evidence, and Impact, McMaster University, Hamilton, ON, Canada

*Corresponding author:*

Correspondence: Jay Park

1280 Main St W, Hamilton, ON L8S 4L8

Email: parkj136@mcmaster.ca

*Co-author details:*

- Rebecca K Metcalfe: rebecca@coreclinicalsciences.com
- Antonio Remiro-Azócar: aazw@novonordisk.com
- Quang Vuong: quang@coreclinicalsciences.com
- Anders Gorst-Rasmussen: agtr@novonordisk.com
- Oliver Keene: oliverkeene405@gmail.com
- Shomoita Alam: shomoita@coreclinicalsciences.com

**Abstract**

The ICH E9(R1) addendum provides guidelines on accounting for intercurrent events in clinical trials using the estimands framework. However, there has been limited attention to the estimands framework for meta-analysis. Using treatment switching, a well-known intercurrent event that occurs frequently in oncology, we conducted a simulation study to explore the bias introduced by pooling together estimates targeting different estimands in a meta-analysis of randomized clinical trials (RCTs) that allowed treatment switching. We simulated overall survival data of a collection of RCTs that allowed patients in the control group to switch to the intervention treatment after disease progression under fixed-effects and random-effects models. For each RCT, we calculated effect estimates for a treatment policy estimand that ignored treatment switching, and a hypothetical estimand that accounted for treatment switching either by fitting rank-preserving structural failure time models or by censoring switchers. Then, we performed random-effects and fixed-effects meta-analyses to pool together RCT effect estimates while varying the proportions of trials providing treatment policy and hypothetical effect estimates. We compared the results of meta-analyses that pooled different types of effect estimates with those that pooled only treatment policy or hypothetical estimates. We found that pooling estimates targeting different estimands results in pooled estimators that do not target any estimand of interest, and that pooling estimates of varying estimands can generate misleading results, even under a random-effects model. Adopting the estimands framework for meta-analysis may improve alignment between meta-analytic results and the clinical research question of interest.

## HIGHLIGHTS

*What is already known:*

The ICH E9(R1) addendum stresses the importance of clearly specifying the estimand of interest in randomized clinical trials with respect to intercurrent events but lacks guidance on how the estimands framework affects meta-analyses.

*What is new:*

We investigated the bias and coverage of treatment effect estimators when estimates from trials targeting estimands with different intercurrent event strategies are pooled in a meta-analysis via a simulation study.

*Potential impact for Research Synthesis Methods readers:*

When conducting meta-analyses, it is important to specify the target estimands of interest and consider an analytical plan that can account for trial-level estimates of different estimands, including their strategies for handling relevant intercurrent events to ensure robust evidence synthesis. Our study illustrates that even a random-effects model cannot handle heterogeneity arising from different estimands in the context of treatment switching. Given that different studies may report estimates targeting different estimands and/or may use different analysis strategies for intercurrent events, there will be an increasing importance in conducting individual patient data meta-analyses rather than ones based on summary statistics. More work is needed to develop meta-analytical methodologies which can account for different estimands in the evidence base.

## 1. INTRODUCTION

In 2019, the International Council for Harmonisation of Technical Requirements for Pharmaceuticals for Human Use (ICH) released an addendum on Estimands and Sensitivity Analysis in Clinical Trials (i.e., the ICH E9(R1) addendum) to highlight the importance of estimands as a way to align the planning, analysis, and interpretation of clinical trials.[1] Notably, the ICH E9(R1) addendum highlights the importance of clearly specifying, during individual trial planning, post-randomization events (also called intercurrent events) that may affect the interpretation of clinical trial outcomes, and strategies to handle these events.[1] Following the addendum's publication, several communications have emphasized the importance of estimands for the design and analysis of clinical trials.[2–9] However, despite mention of implications for meta-analysis in the addendum, there has been limited discussion of the implications of the estimands framework, and particularly intercurrent events, for evidence synthesis.

In oncology, treatment switching is a well-known and common intercurrent event.[10, 11] Here, patients can discontinue their assigned treatment and start an alternative treatment. Control patients are often allowed to switch to the experimental treatment arm after disease progression.[11] It has been reported that the rate of treatment switching is as high as 88% in some oncology trials.[12]

Historically, two common analytical approaches for clinical trials are intention-to-treat and per-protocol analyses.[13] Key principles of intention-to-treat analyses involve analyzing all data from enrolled participants by their randomized allocation, as opposed to the treatment they actually received.[14] Per-protocol analyses, on the other hand, include only the subset of participants that adhered to the trial protocol without major protocol violations. For treatment switching, the intention-to-treat analysis would ignore the treatment switching and target the treatment effects of experimental therapy as randomized, regardless of whether participants switched treatments during the study. This is analogous to the treatment policy estimand under the estimands framework. Per-protocol analyses of cancer trials with treatment switching do not translate to one single target estimand as trial protocols may permit treatment switching based on different criteria, leading to estimands that reflect different treatment plans. The ICH E9(R1) addendum supports analyses that align with estimands that differ from treatment policy. For instance, a hypothetical estimand, where one hypothesizes a scenario in which the intercurrent event would not have taken place, may be more relevant depending on the question of interest.[1] Other hypothetical estimands corresponding to different scenarios may also be specified to better match clinical scenarios observed in practice.[15]

For time-to-event outcomes, commonly used in oncology, there are several existing estimation methods for an estimand of a hypothetical strategy for treatment switching. Simple methods, such as censoring switchers at the point of switch or excluding them entirely from the analysis, can be prone to selection bias as switching is likely to be associated with prognosis.[16] In 2014, the National Institute for Health and Care Excellence (NICE)'s Decision Support Unit published a Technical Support Document (TSD) describing potential analytical methods for situations where control patients in a randomized clinical trial (RCT) are allowed to switch onto the experimental treatment (TSD 16).[17] These methods, including rank-preserving structural failure time modelling (RPSFTM), inverse probability of censoring weighting (IPCW), iterative parameter estimation (IPE), and two-stage estimation (TSE), may be less susceptible to selection bias given other assumptions are satisfied. In April 2024, NICE updated the TSD to discuss broader treatment switching situations where the experimental treatment patients could switch to the control arm, or patients randomized to either trial arm could switch onto treatments not studied in the trial.[16]

Despite the existence of these methods to account for treatment switching, adoption in the analysis of individual RCTs has been limited.[10] A systematic literature review conducted by Sullivan et al., 2020 noted inadequate reporting of methods to account for treatment switching in the analysis of individual RCTs.[10] The two most common analytical strategies for handling treatment switching included: 1) ignoring treatment switching as an intercurrent event under the treatment policy estimand (analogous to an intention-to-treat analysis); and 2) censoring patients at the point of treatment switching under the hypothetical estimand.[18]

To examine methods for addressing treatment switching in evidence synthesis, we, as a separate study, conducted a systematic literature review (PROSPERO: CRD42023487365) of oncology meta-analyses published in the Cochrane Library.[18] The Cochrane Library is widely recognized as the gold standard for evidence synthesis.[19] Similar to inadequate reporting practices in analyses of RCTs,[4] current meta-analytical practices are unsatisfactory for treatment switching as an intercurrent event.

The Cochrane Library provides guidance for incorporating crossover trials in meta-analyses,[20] but this is not an appropriate framework to address treatment switching because switching events in a crossover trial are not prognostic and only relate to the assignment of interventions. In contrast, switching events in oncology trials may be prognostic and depend on properties of the interventions. For evidence synthesis, no meta-analyses reviewed accounted for different trial-level analytical approaches for treatment switching when pooling observed hazard ratios. In other words, estimates targeting different estimands were pooled in meta-analyses.

The objective of this work was to explore the impact of pooling trial estimates targeting differing estimands in meta-analysis. We conducted a simulation study to assess the potential bias associated with current meta-analytical practices that ignore different estimands from individual trials and when control patients are allowed to switch to the treatment arm after disease progression. We compared meta-analyses that pool effect estimates of varying proportions of treatment policy and hypothetical estimands to meta-analyses that pool estimates of only treatment policy or only hypothetical estimands. We chose to estimate the hypothetical estimand using RPSFTM in our main simulations and censoring at the time of treatment switching in our supplementary simulations. RPSFTM was used as it is a recommended method to adjust for treatment switching and it does not require covariate information.[17,21] Censoring was selected as our previous review indicated that this was the most common analytical approach used to handle treatment switching in clinical trials. We focused on the treatment policy estimand as the target meta-analytical estimand because treatment policy is often the estimand preferred by Health Technology Assessment (HTA) bodies.[22] Based on the treatment policy estimand, we estimated the bias and the coverage of the 95% confidence intervals from a pairwise meta-analysis of RCTs that employed different analytical strategies for treatment switching (i.e., treatment policy and hypothetical estimands). Simulated RCTs included in the pairwise meta-analysis estimated overall survival (OS) via hazard ratios (HRs).

In Section 2, we describe our simulation methods in accordance with the ADEMP (Aims, Data-generating mechanisms, Estimands, Methods, and Performance measures) framework for pre-specification of simulation studies.[23] We report our simulation results in Section 3. A discussion then follows (Section 4) along with concluding remarks (Section 5).

## 2. METHODS

This simulation study was performed using a pre-specified ADEMP protocol developed before execution of the simulations. The **A**ims, **D**ata generating mechanisms, **E**stimands, **M**ethods, and **P**erformance measures are described below.

### *2.1 Aims*

We aimed to calculate the bias and coverage of meta-analytical estimators that pool estimates of the treatment policy and hypothetical estimands in varying proportions with respect to the treatment policy estimand.

### *2.2 Data generating mechanisms*

#### ***2.2.1 Illness-death model***

For simulation of an individual trial, we used a three-state irreversible illness-death model. The illness-death model uses a flexible multi-state framework to jointly model progression-free survival (PFS) and OS.[24] There were three states: initial state (state 0); progressed state (state 1); and death (state 2). All subjects started in the initial state. The transition from initial state to progression was governed by the transition hazard $h_{01}(t)$; the transition from progression to death was governed by $h_{12}(t)$; and the transition from initial state to death was governed by $h_{02}(t)$.

#### ***2.2.2 Individual trial simulations based on a real-world trial***

We simulated the PFS and OS times such that their Kaplan-Meier (KM) curves were visually similar to the published KM curves from the PROFound study (NCT02987543).[25–27] The PROFound study was a phase III, open-label RCT in metastatic castration-resistant prostate cancer (mCRPC) that evaluated an oral poly(ADP-ribose) polymerase inhibitor (PARPi). In this RCT, participants randomized to the control arm were allowed to switch treatments after disease progression. A follow-up publication on PROFound by Evans et al., 2021[27] compared various methods to account for treatment switching. We visually inspected our simulated KM curves by contrasting them against pseudo-individual patient data (pseudo-IPD) based on digitized KM curves from PROFound.[28]

For survival times in the treatment group, we tuned the piecewise-constant $h_{01}$ and $h_{02}$ hazards such that the KM curves of the simulated time from randomization to progression and death each had a similar shape to the published KM curves of the PFS and OS of the treatment group reported in the PROFound study. We note that it was specifically the simulated time from randomization to progression and death, not the simulated PFS and OS, that was tuned to match the published curves; this was a deliberate simplification, as it is difficult to derive transition hazards in an illness-death model to obtain a given hazard function for OS. The $h_{01}$ hazard was further tuned on trial-and-error basis to achieve progression proportions of approximately 50% and 75% in our simulations. The $h_{12}$ hazard, which assumed a piecewise-constant form with one change-point at $t = 12$, was tuned such that the median post-progression survival of the simulated data was similar to the difference between the median PFS and OS in the treatment group in the PROFound study. For the control group, we multiplied the $h_{01}$ and $h_{02}$ hazards by the reciprocal of the specified transition hazard ratio $\beta$.

To simulate the effects of switching from the control group to the treatment group, we assumed that all progressors in the control group would switch to the treatment group at the time of progression. Thus, the progression proportions of 50% and 75% reflect switching proportions of 50% and 75%. We chose these switching proportions because the switching proportion in the control arm of the PROFound study was about 80%,[26] and we sought to demonstrate the behavior of meta-analytical estimators for moderate to frequent switching. We assumed that the treatment effect would wane after progression. The magnitude of treatment effect waning was obtained from a review conducted by Kuo, Weng & Lie (2022)[29] that compared the OS from initiation of therapy versus post-progression overall survival. To reflect this waning, we applied a weighted average of hazards where switchers are assumed to experience a reduced hazard of 0.66. This weighted average is then expressed as a multiplicative factor applied to the post-progression hazard among switchers to yield an appropriate population-level average hazard. More simply, the $h_{12}$ hazard of the control group was multiplied by $\frac{0.66(1-\beta)+\beta}{\beta}$.

For each trial, we set: uniform recruitment rate with recruitment to finish in 24 months; 5% random drop-out rates; and an overall trial duration of 48 months to induce administrative censoring. We considered no other intercurrent events. For the analysis of individual trials, we used a simple (univariable) Cox proportional hazards regression of OS on treatment to obtain hazard ratio estimates for the treatment effect on OS.

We considered a total of 12 scenarios with varying treatment effects reflected by different HRs of 0.60, 0.80, and 1.00 assumed for the transition hazards of the illness-death model; switching proportions of 50% and 75%; and unequal (2:1) and equal (1:1) allocations (treatment:control ratio) (Table 1). An unequal allocation ratio of 2:1 was used to match the allocation ratio used in the PROFound trial.[25–27]

### *2.2.3 Meta-analysis*

For each replicate in a given simulation scenario, we simulated $n$ individual trials using the data-generating mechanism above to be pooled in meta-analyses. We used the same transition HR for all trials in each replicate, thus assuming a fixed-effects model for data generation. We specified that each meta-analysis consisted of $n = 8$ RCTs based on other simulation studies of meta-analyses.[30,31] The sample size of each trial was randomly chosen to be 250, 300, or 350 with equal probability. These possible sample sizes were chosen to be similar to the sample size of the PROFound trial, which was 245.[26] For each scenario, we generated 10,000 replicates (10,000 meta-analyses of 8 trials each, corresponding to a total of 80,000 simulated trials).

To ensure robustness, we repeated the entire simulation process using a random-effects model for data generation. Here, for $n$ trials in a replicate under a scenario where the transition HR was $\beta$, we first sampled $n$ study-specific log transition HRs $\log\beta_1, \ldots, \log\beta_n$ from $N(\log\beta, \tau^2)$ for a pre-selected $\tau^2$ of 0.03. We selected this value of $\tau^2$ because it was the median of the reported values of $\tau^2$ for treatment effects on OS, in the log HR scale, in our review of meta-analyses in the Cochrane Library.[18] Then the individual trials were generated as above using each $\beta_i$ as the specified transition HR. We used the same number of trials, sample size and number of replicates as in the main simulation.

*2.3 Estimands*

We primarily considered a treatment policy estimand as our target meta-analytical estimand. Under the treatment policy estimand, treatment switching for the control patients after disease progression would be ignored for the comparison of OS. The target of our simulations was to quantify the bias in HRs of OS estimated from pairwise meta-analyses pooling individual RCT results reflecting varying proportions of estimates targeting treatment policy and hypothetical estimands at the level of individual trials.

Under an illness-death model, the proportional hazards assumption is violated even when the transition hazards satisfy the proportional hazards assumption with respect to treatment.[24] Simulated data must be used to approximate the true value. An exception is the null hypothesis-like scenario with a prespecified transition HR of 1, where the treatment policy OS HR estimand is also equal to 1. For our scenarios with the transition HRs of 0.60 and 0.80, we simulated a large trial with a sample size of 1,000,000. The estimated treatment policy OS HR value using the trial-specific analytical method for the treatment policy estimand in Section 2.4.1 was used as the "true" value of this estimand. Upon informal inspection, the value of the true HR was stable up to two decimal places over repeated simulations.

Different data-generating mechanisms have different implications for the estimands. Our main fixed-effects simulation assumes that there is one single treatment policy OS HR estimand at the individual study and meta-analytical levels. There is no heterogeneity between true treatment effect estimands across studies beyond that induced by different intercurrent event strategies. Conversely, our random-effects simulation assumes there are distributions of heterogeneous treatment policy estimands cross trials. Such heterogeneity could be due to unexplained factors in the high-level distinction between different estimand types, e.g. details around intercurrent event strategy, population, treatment implementation or outcome. The true meta-analytical OS HR in the random-effects setting is characterized by the mean $\beta$ of the underlying normal distribution of transition log HRs.

*2.4 Methods*

***2.4.1 Estimation of trial-level treatment policy and hypothetical estimands***

For each simulated trial, we estimated HRs targeting the treatment policy and hypothetical estimands. To estimate the treatment policy estimand, the OS time was compared between the control and experimental groups according to initial treatment assignment, with the HR as the population-level summary measure. The OS time of control patients who switched contains the survival period they spent receiving the experimental treatment. We obtained estimates by fitting a simple (univariable) Cox proportional hazards regression with OS as the outcome and treatment as the only predictor. From the fitted model, we extracted the estimated log HR under each intercurrent event strategy, corresponding to the estimated treatment coefficient, and its model-based nominal standard error.

To estimate the hypothetical estimand, we used RPSFTMs and censoring at the time of switching in separate simulations. We consider the simulations involving RPSFTMs to be our main simulations, while the simulations involving censoring switchers are the supplementary simulations.

Let $T_0$ and $T_1$ be the amount of time a patient spends in the control and treatment groups, respectively. The RPSFTM assumes that the counterfactual survival time of a patient if they were always in the control group, $U$, satisfies

$$U = T_0 + e^{\psi} T_1$$

for an acceleration factor $e^{\psi}$.[21] We estimated $\psi$ using g-estimation as implemented in the R package rpsftm.[32] Then, with the estimator $\hat{\psi}$, survival times of switchers in the control group were adjusted by $T_0 + e^{\hat{\psi}} T_1$; survival times of all other patients were unadjusted. Recensoring was applied by multiplying administrative censoring times of patients in the control group by $e^{\hat{\psi}}$ and updating censoring indicators accordingly; censoring times in the treatment group were unadjusted.[21,32] A Cox proportional hazards model was fit to the new survival times to extract an estimate of the log HR. The standard error was calculated so that this analysis has the same $p$-value as the analysis for treatment policy estimand.[21] The details of the supplementary analysis censoring switchers at the time of switch are provided in the Appendix Section 1.

### *2.4.2 Meta-analytical synthesis*

For each collection of eight trials, we performed a random-effects meta-analysis using the inverse variance method to synthesize the estimated trial-specific treatment effects.[33] The meta-analysis was done on the log-scale, where the log HR estimates were pooled with the inverse of their estimated variances as weights, and the pooled estimate was back-transformed to the HR scale. The standard error of the pooled log HR estimate was computed assuming independence of trials, with the between-study variance estimated using restricted maximum likelihood (REML). 95% confidence intervals were also computed on the log scale with

$$\widehat{log(HR)} \pm 1.96\widehat{SE}\left(log(HR)\right)$$

and then back-transformed to the HR scale. We calculated the pooled HR estimates, with different proportions of RCTs targeting treatment policy and hypothetical estimands being pooled in a given meta-analysis. In each meta-analysis, we varied the proportion of RCTs with a treatment policy estimand at 0, 0.25, 0.50, 0.75, and 1.00. This in turn meant that the proportion of RCTs with a hypothetical estimand in each meta-analysis varied at 1.00, 0.75, 0.50, 0.25, and 0, respectively.

As a sensitivity analysis, we performed a fixed-effects meta-analysis for trials in each replicate using the inverse variance method.[33] This analysis was performed largely similarly to the random-effects meta-analysis, but the between-study variance was set to 0. We also varied the proportion of RCTs with a treatment policy or hypothetical estimand in the same way as above.

The simulations we conducted are summarized in Tables 1 and 2. In total, 12 simulation scenarios were considered, varying the transition HR, switching proportion, and allocation ratio. Within each scenario we conducted six settings, varying the hypothetical estimand estimator (either RPSFTMs or censoring switchers), fixed- or random-effects data generation, and fixed- or random-effects meta-analytical synthesis. We considered the simulations with fixed-effects data-generating mechanisms and random-effects meta-analysis estimation to be the primary simulations for our main RPSFTM and supplementary censoring switchers estimators, and all other simulations to be sensitivity analyses.

*2.5 Performance Measures*

Our performance measures of interest were the bias and 95% confidence interval coverage of the pooled estimators, constructed using varying proportions of estimates targeting hypothetical and treatment policy estimands. We calculated the bias and coverage with respect to the treatment policy estimand. The performance measures were calculated in each scenario for the meta-analytical estimators specified in the simulations.

Let $\hat{\delta}_j$ and $\widehat{se}(\hat{\delta}_j)$ be the pooled HR estimate and its estimated standard error for the $j = 1, \ldots, N$th set of $n$ simulated trials. For clarity, $N = 10{,}000$ and $n = 8$. With $\delta$ being the true value of an estimand, the bias is estimated with

$$\widehat{Bias} = \frac{1}{N}\sum_{j=1}^{N} \hat{\delta}_j - \delta,$$

and the coverage is estimated with

$$\widehat{Coverage} = \frac{1}{N}\sum_{j=1}^{N} I\left(\hat{\delta}_j - 1.96\widehat{se}(\hat{\delta}_j) \leq \delta \leq \hat{\delta}_j + 1.96\widehat{se}(\hat{\delta}_j)\right).$$

We specified the calculation of the true value of this estimand in Section 2.3. Note that the absolute bias was reported on the HR scale instead of on the log HR scale.

We quantified the uncertainty of the performance measures using Monte Carlo standard errors. We calculated these as follows. The standard error of the bias was calculated as

$$\widehat{SE}_{Bias} = \frac{sd(\hat{\delta}_j)}{\sqrt{N}},$$

and the standard error of the coverage was calculated as

$$\widehat{SE}_{Coverage} = \sqrt{\frac{\widehat{Coverage}(1 - \widehat{Coverage})}{N}}.$$

*2.6 Software*

We performed our simulation study using R software version 4.3.2.[34] We used the R packages survival[35] to fit Cox proportional hazards models and estimate the hazard ratio for each individual study, rpsftm[32] to fit RPSFTMs, meta[36] to perform the meta-analyses, and simIDM[37] to simulate the data. The results were visualized using the ggplot2 package[38] and tabulated using the flextable[39] and officer[40] R packages. This manuscript was prepared using Quarto via RStudio.[41,42]

## 3. RESULTS

We present the results for the random-effects meta-analytical estimators with fixed-effects data generation integrating trial-level estimates reflecting treatment policy and hypothetical estimands. The simulation results of 12 different scenarios explored in this study are organized by their specified transition HRs of our illness-death model. The base case of our simulation involved scenarios with the specified transition HR of 0.60 and varying allocation ratios and treatment switching rates. The results of other simulations are presented in the Appendix.

### *3.1 Main simulations*

#### ***3.1.1 Base case scenarios under assumed HR of 0.60 for the transition hazards of the illness-death model***

Figure 1 presents density plots showing the distribution of point estimates of the HRs from different meta-analytical estimators under the specified transition HR of 0.60. Table 3 shows the average of the point estimates, lower and upper bounds of the averaged 95% CIs, and the calculated bias and coverage of different estimators under the specified transition HR of 0.60. The Monte Carlo standard errors of all performance measures were less than 0.005.

On average, pooling purely hypothetical estimates produced stronger treatment effects than pooling purely treatment policy estimates. Here, "pure" refers to the meta-analytic estimator obtained by pooling trial-level estimates under a given estimand strategy (hypothetical or treatment policy), and should be distinguished from the "true" estimand, which is defined by the data-generating mechanism. This was true across allocation ratios and control arm treatment switching rates. For instance, the average treatment effect pooling purely hypothetical estimates with unequal (2:1) allocation and 75% switching rate for the control arm was 0.41 (averaged 95% CI: 0.34, 0.51) compared to 0.66 for pooling purely treatment policy estimates (averaged 95% CI: 0.60, 0.73). The pure treatment policy pooling strategy generally yielded smaller treatment effect estimates with the higher treatment switching rate of 75% compared to 50%. On the other hand, the pure hypothetical pooling strategy yielded larger treatment effect estimates under the 75% switching rate compared to the 50% switching rate for both unequal and equal allocations. For a given treatment switching rate, there were also negligible differences between unequal and equal allocation ratios.

With respect to the treatment policy estimand, the bias and coverage of meta-analyses worsened when the proportion of hypothetical estimates included in the pooling increased. In scenarios with unequal allocation and a 75% switching rate, the meta-analytical estimator that pooled 25% treatment policy estimates (75% hypothetical estimates) had a bias of -0.16 (2.5 and 97.5% percentiles: -0.22 and -0.07), whereas the meta-analytical estimator that pooled 75% treatment policy estimates had a smaller bias of -0.03 (2.5 and 97.5% percentiles: -0.10, 0.04). Coverage with respect to the treatment policy estimand decreased as the meta-analytical estimators included a larger proportion of trials reporting hypothetical estimates.

#### ***3.1.2 Alternate scenarios under assumed HRs of 0.80 and 1.00 for the transition hazards of the illness-death model***

The density plots of point estimates of the HRs estimated under assumed transition HRs of 0.80 and 1.00 (null scenario) are shown in Figure 2 and Figure 3, respectively. Performance in terms of bias and coverage under these specified transition HRs is shown in Table 4 and Table 5. Similar to scenarios with the specified transition HR of 0.60, the Monte Carlo standard errors of all performance measures were less than 0.005. The findings of the simulations under the specified transition HR of 0.80 are similar to the findings of the scenarios under the specified transition HR of 0.60. We saw stronger treatment effects estimated from the meta-analytical estimator pooling purely hypothetical estimates than that pooling purely treatment policy estimates, across all allocation ratios and treatment switching rates. With respect to the treatment policy estimand, both bias and coverage worsened as the meta-analytical estimators pooled a larger proportion of estimates of the hypothetical estimand.

For the null scenarios (transition HR of 1.00, corresponding to an OS HR of 1.00), the average estimated treatment effects for OS of different meta-analytical estimators generally were close to 1.00. Here, both the treatment policy and hypothetical OS HR estimands were 1, representing no treatment effect. As a result, there was generally very little bias in different meta-analytical estimators when compared to the treatment policy estimand. The coverage of the pure treatment policy estimator across all allocation ratios and switching rates was 0.96, close to the appropriate nominal value of 0.95. This suggests adequate variance/interval estimation. Conversely, there was notable under-coverage for the pure hypothetical estimator across all allocation ratios and switching rates, with coverage as low as 0.86. This might be due to the ad hoc strategy used to compute standard errors at the trial level leading to overprecision.[21]

### *3.1.3 Sensitivity analyses in main simulations*

The sensitivity analytical results with the fixed-effects meta-analysis as well as random-effects data-generating mechanism for the main simulations with RPSFTM are provided in the Appendix Section 2. With fixed-effects meta-analysis (Simulation 2, Appendix Section 2.1), we saw similar findings as the random-effects meta-analyses with a fixed-effect data-generating mechanism. With respect to the treatment policy estimand, the fixed-effects meta-analytical estimators had similar bias to the random-effects meta-analytical estimators, but the coverage of the fixed-effects meta-analytical estimators that included hypothetical estimators was lower than the corresponding random-effects meta-analytical estimators. This was due to the smaller standard errors estimated by fixed-effects meta-analyses.

Under random-effects data generation (Simulation 3, Appendix Section 2.2), random-effects meta-analytical estimators exhibited similar behaviors when estimators targeting different intercurrent event strategies were pooled. The magnitude of average biases and coverage for each of the respective meta-analytical estimators with reference to the treatment policy estimand were generally similar. However, there were increased variabilities in the biases as noted by the wider range of the difference percentiles, and the coverage of all meta-analytical estimators generally decreased due to the random-effects data generation and the finite number of trials in the meta-analysis.

## *3.2 Supplementary simulations*

To supplement the main simulations where the RPSFTM was used as trial-level estimator for the hypothetical estimand, we performed additional simulations where censoring at the time of treatment switching was used. The results can be found in the Appendix Section 3.

The results showed broadly similar patterns to the main simulations. For transition HRs of 0.60 and 0.80, with respect to the treatment policy estimand, the bias and coverage of meta-analytical estimators worsened as a higher proportion of hypothetical estimators were pooled. However, it is noteworthy that censoring switchers produced smaller effect estimates than RPSFTM. Because of this, the magnitude of the bias was smaller than that in the main simulations. Likewise, the reduction in coverage was less drastic compared to the main simulations. For the transition HR of 1.00, all meta-analytical estimators showed small bias and adequate coverage of at least 0.95.

## 4. DISCUSSION

In this study, we explored how pooling trial-level estimates of treatment policy and hypothetical estimands affects meta-analyses of oncology trials in presence of treatment switching. Using the treatment policy estimand as our target meta-analytical estimand, we specifically explored the quantitative bias associated with pooling HRs of OS under different analytical strategies for treatment switching after disease progression in the patients allocated to the control arm. The bias of the pooled estimator relative to the target estimand and the corresponding coverage of confidence intervals worsened as a greater proportion of hypothetical trial-level estimates were included in the meta-analysis. Our simulations showed that the frequency of the intercurrent event also affects the magnitude of bias. Consistent results were observed across two common analytical strategies of RPSFTM and censoring at the time of treatment switching for estimating trial-level hypothetical estimands.

Our simulations provide quantitative insights into the bias that arises when estimates of different estimands for treatment switching are combined in meta-analyses. We demonstrated that when different estimates are combined naïvely (i.e., without consideration of the differing estimands), meta-analyses produce a pooled estimate that does not reflect any specific target estimand. While our simulations assessed two analytical strategies for treatment switching under the hypothetical estimand, other analytical strategies are possible, such as two-stage estimation approaches and models using inverse probability of censoring weights.[27] Each analytic strategy can yield a treatment effect estimate that differs from another.[27] This may be explained by the fact that different analytical strategies impose different modeling assumptions on the relationship between the outcome and intercurrent events. Indeed, in our main and supplementary simulations, RPSFTMs and censoring switchers yield different estimates of the hypothetical estimand, even though the hypothetical scenario targeted by both strategies was specified to be identical. Latimer and colleagues reported similar results in their investigation of different adjustment methods for treatment switching.[43] Therefore, we expect that pooling hypothetical estimates addressed by different analytical strategies may yield similar trends as those observed when pooling hypothetical estimands with treatment policy estimands.

Meta-analyses are a crucial tool for clinical research. The findings generated from meta-analyses have important implications for clinical practice and policy decisions, including reimbursement and access to potentially life-saving therapies. In this study, the magnitude of the bias induced by pooling estimates from different estimands was large enough to impact cost-effectiveness estimates, such as those used by HTA bodies to make reimbursement decisions. For example, sensitivity analyses conducted as part of the evidence package for NICE's appraisal of pazopanib found that changes in the point estimate of HR for OS from 0.563 to 0.636 resulting from different strategies for treatment switching moved the treatment from cost-effective to cost-ineffective.[44] Indeed, survival parameters are often among the most influential variables in cost-effectiveness analyses of oncology therapies.[45–49] Our findings suggest that naïve pooling of trial estimates when different strategies are used for intercurrent events, especially when they occur as frequently as treatment switching, may be difficult to interpret. Naïve pooling of these different trial results could potentially result in life-saving cancer therapies being deemed ineffective (or less effective) and not cost-effective, or conversely, ineffective therapies being deemed effective (or more effective).

In evidence synthesis, we often use the PICO (population, intervention, comparator, and outcome) framework to translate policy questions to research questions that then determine the scope of systematic literature reviews and meta-analyses.[50] Broad PICO statements are often used to capture a large body of literature that can reflect the totality of scientific evidence for clinical and policy decision making. Compared to the PICO framework, an important distinction of the estimand framework is specificity in relevant intercurrent events that could change the interpretation of trial results and their respective analytical strategies.[51] However, this distinction is missing from current guidance for meta-analysis. For example, the Cochrane Handbook for Systematic Reviews of Interventions does not provide guidance on how intercurrent events should be considered when conducting systematic reviews.[20] The recently published Methods Guide for Health Technology Assessment by Canada's Drug Agency (CDA-AMC) explicitly calls for identification of different estimands and intercurrent events for individual clinical trials included in the evidence base.[52] However, this document still lacks guidance on pooling studies that have different intercurrent events of interest and analytical strategies.[52]

The central themes of the ICH E9(R1) addendum are the importance of carefully considering relevant intercurrent events and clearly describing the treatment effect that is to be estimated for correct interpretation of trial results. While discussion of the addendum has largely pertained to individual RCTs themselves, these insights are equally relevant for evidence synthesis methods,[51] and guidance on these methods should explicitly describe the role of intercurrent events in systematic reviews. By improving transparency around the handling of important intercurrent events, the estimands framework may improve how meta-analyses are designed, conducted, and reported.

Strengthened alignment with the estimands framework would likely bring important changes. As different studies may report estimates targeting different estimands and/or may use different analytical strategies to handle intercurrent events, there would be an increasing importance in conducting meta-analyses based on individual patient data rather than summary statistics. Pharmaceutical companies and academic research groups are increasingly allowing access to the data from their trials making such meta-analyses more feasible.[53] The divergence between treatment policy and hypothetical estimands increases with the rate of treatment switching. The importance of intercurrent events to meta-analysis depends on their frequency. Using the estimands framework may help researchers identify which intercurrent events are most likely to alter the interpretation of the study treatment effect based on their anticipated frequency. Even so, requiring more consistent handling of common intercurrent events across studies may result in sparse evidence bases that consist of fewer trials. This has important implications for network meta-analyses (NMAs): a sparser evidence base may result in disconnected networks limiting feasibility.[54] Regardless, NMAs conducted with different treatment effects estimated under different strategies for relevant intercurrent events should proceed with caution, as bias can propagate through the evidence network, impacting the accuracy not just of one treatment comparison, as in pairwise meta-analysis, but of multiple treatment comparisons.[55,56] The development of new meta-analytic methods to handle heterogeneity in pooled estimands would counteract this challenge while retaining the increased specificity offered by the estimands framework.

It is important to consider our findings in the context of our study's limitations. Our study is a simulation-based study and lacks a real case study. However, a simulation study is more appropriate to demonstrate the bias of meta-analytical estimators than a real case study because a simulation study allows us to have knowledge of the true underlying model and parameters, and we designed our simulations based on a real case study, namely the PROFound study.[26] Our simulation study is narrow in scope. We considered a limited number of scenarios in terms of the trial sample size, the number of studies in a meta-analysis, and the switching proportions that were chosen based on the existing literature such that our simulation mimics meta-analytical approaches used in practice.[26,30,31] We assumed that studies targeting the hypothetical estimand used the same analytical strategy to estimate it, and that the specified hypothetical estimand was identical across all these studies.

Most importantly, we only considered treatment switching from the control arm to the experimental treatment arm due to disease progression. There are other forms of treatment switching where patients randomly assigned to the experimental treatment arm could switch to the control arm or patients can switch onto other treatments not studied in the trial.[16] In practice, a clinical trial may allow treatment switching for many reasons other than disease progression (e.g., patient intolerability, lack of efficacy, preference, and clinical discretion). Furthermore, we assumed that after progression, all participants in the control arm received the experimental treatment. This is similar to the PROFound study, [26] as well as other studies,[12] where the vast majority of control participants switched to the experimental treatment after progression. Although in these studies not every participant switched to the experimental treatment, it is unlikely that this assumption would alter our primary finding that pooling trial estimates targeting two different estimands yields meta-analytic estimates that may not reflect either target estimand. In addition to treatment switching, there are other intercurrent events that were not considered in our simulations. It is likely that less common intercurrent events would introduce less bias into meta-analytical estimators. Regardless, our findings highlight the need for clarity in the target estimand for meta-analysis. It is likely that pooling evidence reflecting estimates of different trial-level estimands may produce biases in the meta-analysis, especially when the intercurrent events of interest occur in high frequency.

### *4.1 Implications for future research*

We have identified several directions for future research. Future simulations may explore a broader range of scenarios, as well as the case where trials targeting hypothetical estimands use different analytical strategies. Our findings show that the estimands framework is highly relevant for evidence synthesis, but discussion of the role of estimands for evidence synthesis has been limited, in particular by non-statisticians. The importance of transparent reporting at the level of individual trials to enable high-quality systematic reviews and meta-analyses cannot be understated. Lee and Torres have proposed reporting guidelines specifically to address challenges of treatment switching.[57] For evidence synthesis of time-to-event outcomes, it is a common data extraction practice to digitize the published KM curves to create pseudo-IPD. Different censoring mechanisms will produce different KM curves, but a previous assessment showed that for many trials, it is difficult to understand the target estimand that is being estimated.[4,18] Of particular note, available KM curves are often limited to the primary analysis that may differ from the target estimand of the meta-analysis. Importantly, this work adds to prior research showing analytic strategies targeting the same estimand can yield different estimates even when model assumptions are met. Further work is needed to determine the contexts in which different analytic strategies, such as two-stage estimation and modelling using inverse probability of censoring weights, are optimal. More work is needed to develop methods that can account for different estimands and analytical strategies for intercurrent events. For a given outcome, it may be possible that treatment effects estimated for different estimands may be combined and synthesized through multivariate normal random-effects meta-analysis.[51,58,59,60] It might be also possible that multi-state network meta-analysis methods for progression and survival data, or illness-death models, may be adapted to handle different estimands.[24,60]

## 5. CONCLUSION

Our study shows that naive pooling of treatment effects estimated under different strategies for treatment switching can produce biased results relative to the target estimand of the meta-analysis. While our study is limited to time-to-event analysis and treatment switching, our findings point to potential challenges in pooling estimates targeting estimands with different intercurrent event strategies in aggregate-level meta-analyses. Having broad research questions can result in a larger evidence base; however, pooling a broad set of studies with treatment effects estimated using different strategies for frequent intercurrent events may lead to misleading results and important consequences for HTA decision making. Adopting the estimands framework for evidence synthesis can result in more relevant estimates of treatment effects that better reflect the clinical questions of interest to both health practitioners and policy decision-makers.

### Author Contributions

Conceptualisation, JJHP; methodology, RKM, ARA, QV, JJHP; software, QV and JJHP; validation, QV and JJHP; formal analysis, ARA, QV; investigation, JJHP; resources, JJHP; data curation, QV; writing—original draft preparation, RKM, QV, and JJHP; writing—review and editing, RKM, QV, ARA, AGR, OK, SA and JJHP; visualization, QV, RKM, and JJHP; supervision, JJHP; project administration, QV and RKM; funding acquisition, JJHP.

All authors have read and agreed to the published version of the manuscript.

### Acknowledgements

Open access publishing facilitated by McMaster University, as part of the Wiley Hybrid Journals - McMaster University agreement via CRKN (Canadian Research Knowledge Network).

We thank Richard Yan for their assistance in coding.

**Funding Information**

There was no funding for this study.

**Conflict of interest statement**

The authors have no other conflicts of interest to disclose.

**Data Availability Statement**

The datasets generated and/or analysed during the current study, in addition to the code to replicate the simulation study in its entirety, will be made available in a public GitHub repository after peer review.

**Statements Relating to Our Ethics and Integrity Policies**

- Ethics approval statement: Not applicable
- Patient consent statement: Not applicable
- Permission to reproduce material from other sources: Not applicable.
- Clinical trial registration: Not applicable

## Tables

*Table 1: Simulation scenarios*

| True transition HR[a] | Switching Proportion | Allocation Ratio |
|---|---|---|
| 0.6 | 0.75 | 2:1 |
| 0.8 | 0.75 | 2:1 |
| 1.0 | 0.75 | 2:1 |
| 0.6 | 0.50 | 2:1 |
| 0.8 | 0.50 | 2:1 |
| 1.0 | 0.50 | 2:1 |
| 0.6 | 0.75 | 1:1 |
| 0.8 | 0.75 | 1:1 |
| 1.0 | 0.75 | 1:1 |
| 0.6 | 0.50 | 1:1 |
| 0.8 | 0.50 | 1:1 |
| 1.0 | 0.50 | 1:1 |

**[a] True transition HR refers to the assumed hazards from one stage to another in our three-state irreversible illness-death model.**

*Table 2: Summary of simulation settings within each scenario*

| Simulation | Meta-analytical data-generating mechanism | Analytical strategy for hypothetical estimand | Meta-analysis model |
|---|---|---|---|
| **Main simulations** | | | |
| Primary | Fixed-effects | RPSFTM[a] | Random-effects |
| Sensitivity | Fixed-effects | RPSFTM | Fixed-effects |
| Sensitivity | Random-effects | RPSFTM | Random-effects |
| **Supplementary simulations** | | | |
| Primary | Fixed-effects | Censoring switchers | Random-effects |

| Sensitivity | Fixed-effects | Censoring switchers | Fixed-effects |
|---|---|---|---|
| Sensitivity | Random-effects | Censoring switchers | Random-effects |

**[a]RPSFTM: rank-preserving structural failure time model**

*Table 3: Averages of pooled treatment effect estimates and comparison against treatment policy estimand under an assumed HR of 0.60 for the transition hazards of the illness-death model in the simulation with fixed-effects data-generating mechanism, random-effects meta-analysis, and rank-preserving structural failure time models*

| **Scenarios** | **Estimators** | **Estimated treatment effects: HR (averaged 95% CI)** | **Bias (2.5%, 97.5% difference percentiles)** | **Coverage** |
|---|---|---|---|---|
| **75% switching rate for control arm** | | | **Comparison against treatment policy estimand (True HR = 0.66)** | |
| 2:1 allocation, 75% switching rate for control arm[a] | Pure HE (100%) | 0.41 (0.34, 0.51) | -0.25 (-0.30, -0.17) | 0.01 |
| | Mixed TPE (25%) and HE (75%) | 0.50 (0.40, 0.63) | -0.16 (-0.22, -0.07) | 0.27 |
| | Mixed TPE (50%) and HE (50%) | 0.58 (0.48, 0.69) | -0.08 (-0.16, 0.00) | 0.77 |
| | Mixed TPE (75%) and HE (25%) | 0.63 (0.56, 0.72) | -0.03 (-0.10, 0.04) | 0.93 |
| | Pure TPE (100%) | 0.66 (0.60, 0.73) | -0.00 (-0.06, 0.06) | 0.96 |
| 1:1 allocation, 75% switching rate for control arm | Pure HE (100%) | 0.41 (0.33, 0.50) | -0.25 (-0.31, -0.18) | 0.00 |
| | Mixed TPE (25%) and HE (75%) | 0.50 (0.39, 0.63) | -0.16 (-0.23, -0.08) | 0.22 |
| | Mixed TPE (50%) and HE (50%) | 0.57 (0.48, 0.69) | -0.09 (-0.16, -0.00) | 0.77 |
| | Mixed TPE (75%) and HE (25%) | 0.63 (0.56, 0.72) | -0.03 (-0.09, 0.04) | 0.94 |
| | Pure TPE (100%) | 0.66 (0.60, 0.73) | 0.00 (-0.05, 0.06) | 0.96 |
| **50% switching rate for control arm** | | | **Comparison against treatment policy estimand (True HR = 0.64)** | |
| 2:1 allocation, 50% switching rate for control arm | Pure HE (100%) | 0.53 (0.46, 0.60) | -0.11 (-0.17, -0.05) | 0.18 |
| | Mixed TPE (25%) and HE (75%) | 0.56 (0.49, 0.65) | -0.08 (-0.14, -0.01) | 0.56 |
| | Mixed TPE (50%) and HE (50%) | 0.59 (0.52, 0.67) | -0.05 (-0.11, 0.02) | 0.81 |
| | Mixed TPE (75%) and HE (25%) | 0.62 (0.55, 0.69) | -0.02 (-0.08, 0.04) | 0.93 |
| | Pure TPE (100%) | 0.64 (0.57, 0.71) | -0.00 (-0.06, 0.06) | 0.96 |
| | Pure HE (100%) | 0.52 (0.46, 0.60) | -0.12 (-0.18, -0.05) | 0.14 |
| 1:1 allocation,<br>50% switching rate for control arm | Mixed TPE (25%) and HE (75%) | 0.56 (0.49, 0.64) | -0.08 (-0.14, -0.01) | 0.53 |
| | Mixed TPE (50%) and HE (50%) | 0.59 (0.52, 0.67) | -0.05 (-0.11, 0.02) | 0.82 |
| | Mixed TPE (75%) and HE (25%) | 0.62 (0.55, 0.69) | -0.02 (-0.08, 0.04) | 0.94 |
| | Pure TPE (100%) | 0.64 (0.58, 0.70) | 0.00 (-0.06, 0.06) | 0.96 |

**[a]This table shows estimated treatment effects under an assumed hazard ratio (HR) of 0.60 for the transition hazards of the illness-death model and bias and coverage in comparison to the true treatment policy estimand. Monte Carlo standard errors for all measures are very close to zero**
**Acronyms: CI: Confidence intervals; HE - Hypothetical estimator; HR - Hazard ratio; TPE - Treatment policy estimator.**

*Table 4: Averages of pooled treatment effect estimates and comparison against treatment policy estimand under an assumed HR of 0.80 for the transition hazards of the illness-death model in the simulation with fixed-effects data-generating mechanism, random-effects meta-analysis, and rank-preserving structural failure time models*

| **Scenarios** | **Estimators** | **Estimated treatment effects: HR (averaged 95% CI)** | **Bias (2.5%, 97.5% difference percentiles)** | **Coverage** |
|---|---|---|---|---|
| **75% switching rate for control arm** | | | **Comparison against treatment policy estimand (True HR = 0.84)** | |
| 2:1 allocation, 75% switching rate for control arm[a] | Pure HE (100%) | 0.65 (0.51, 0.83) | -0.19 (-0.31, -0.03) | 0.42 |
| | Mixed TPE (25%) and HE (75%) | 0.74 (0.61, 0.90) | -0.10 (-0.22, 0.04) | 0.77 |
| | Mixed TPE (50%) and HE (50%) | 0.79 (0.69, 0.92) | -0.04 (-0.15, 0.07) | 0.90 |
| | Mixed TPE (75%) and HE (25%) | 0.82 (0.73, 0.93) | -0.02 (-0.10, 0.08) | 0.94 |
| | Pure TPE (100%) | 0.84 (0.76, 0.93) | -0.00 (-0.08, 0.08) | 0.96 |
| 1:1 allocation, 75% switching rate for control arm | Pure HE (100%) | 0.65 (0.52, 0.82) | -0.19 (-0.31, -0.03) | 0.39 |
| | Mixed TPE (25%) and HE (75%) | 0.74 (0.62, 0.90) | -0.10 (-0.21, 0.04) | 0.77 |
| | Mixed TPE (50%) and HE (50%) | 0.80 (0.69, 0.92) | -0.04 (-0.14, 0.06) | 0.91 |
| | Mixed TPE (75%) and HE (25%) | 0.83 (0.74, 0.92) | -0.01 (-0.10, 0.07) | 0.95 |
| | Pure TPE (100%) | 0.84 (0.76, 0.93) | 0.00 (-0.07, 0.08) | 0.96 |
| **50% switching rate for control arm** | | | **Comparison against treatment policy estimand (True HR = 0.83)** | |
| 2:1 allocation, 50% switching rate for control arm | Pure HE (100%) | 0.74 (0.64, 0.87) | -0.08 (-0.18, 0.03) | 0.68 |
| | Mixed TPE (25%) and HE (75%) | 0.77 (0.67, 0.89) | -0.05 (-0.15, 0.05) | 0.83 |
| | Mixed TPE (50%) and HE (50%) | 0.80 (0.70, 0.90) | -0.03 (-0.12, 0.07) | 0.90 |
| | Mixed TPE (75%) and HE (25%) | 0.81 (0.73, 0.91) | -0.01 (-0.09, 0.07) | 0.94 |
| | Pure TPE (100%) | 0.82 (0.74, 0.91) | -0.00 (-0.08, 0.08) | 0.96 |
| | Pure HE (100%) | 0.74 (0.64, 0.86) | -0.08 (-0.18, 0.03) | 0.67 |
| 1:1 allocation,<br>50% switching rate for control arm | Mixed TPE (25%) and HE (75%) | 0.77 (0.68, 0.89) | -0.05 (-0.14, 0.05) | 0.84 |
| | Mixed TPE (50%) and HE (50%) | 0.80 (0.71, 0.90) | -0.03 (-0.11, 0.06) | 0.91 |
| | Mixed TPE (75%) and HE (25%) | 0.81 (0.73, 0.91) | -0.01 (-0.09, 0.07) | 0.95 |
| | Pure TPE (100%) | 0.83 (0.75, 0.91) | 0.00 (-0.07, 0.08) | 0.96 |

**[a]This table shows estimated treatment effects under an assumed hazard ratio (HR) of 0.80 for the transition hazards of the illness-death model and bias and coverage in comparison to the true treatment policy estimand. Monte Carlo standard errors for all measures are very close to zero**
**Acronyms: CI: Confidence intervals; HE - Hypothetical estimator; HR - Hazard ratio; TPE - Treatment policy estimator.**

*Table 5: Averages of pooled treatment effect estimates and comparison against treatment policy estimand under an assumed HR of 1.00 for the transition hazards of the illness-death model in the simulation with fixed-effects data-generating mechanism, random-effects meta-analysis, and rank-preserving structural failure time models*

| **Scenarios** | **Estimators** | **Estimated treatment effects: HR (averaged 95% CI)** | **Bias (2.5%, 97.5% difference percentiles)** | **Coverage** |
|---|---|---|---|---|
| **75% switching rate for control arm** | | | **Comparison against treatment policy estimand (True HR = 1.00)** | |
| 2:1 allocation, 75% switching rate for control arm[a] | Pure HE (100%) | 0.99 (0.77, 1.28) | -0.01 (-0.23, 0.26) | 0.89 |
| | Mixed TPE (25%) and HE (75%) | 0.99 (0.84, 1.19) | -0.01 (-0.16, 0.17) | 0.91 |
| | Mixed TPE (50%) and HE (50%) | 1.00 (0.87, 1.15) | -0.00 (-0.13, 0.13) | 0.93 |
| | Mixed TPE (75%) and HE (25%) | 1.00 (0.89, 1.12) | -0.00 (-0.11, 0.11) | 0.94 |
| | Pure TPE (100%) | 1.00 (0.90, 1.11) | -0.00 (-0.10, 0.10) | 0.96 |
| 1:1 allocation, 75% switching rate for control arm | Pure HE (100%) | 1.00 (0.79, 1.28) | 0.00 (-0.21, 0.26) | 0.90 |
| | Mixed TPE (25%) and HE (75%) | 1.00 (0.85, 1.18) | -0.00 (-0.15, 0.17) | 0.91 |
| | Mixed TPE (50%) and HE (50%) | 1.00 (0.88, 1.14) | -0.00 (-0.12, 0.13) | 0.93 |
| | Mixed TPE (75%) and HE (25%) | 1.00 (0.89, 1.12) | 0.00 (-0.10, 0.11) | 0.95 |
| | Pure TPE (100%) | 1.00 (0.91, 1.11) | 0.00 (-0.09, 0.10) | 0.96 |
| **50% switching rate for control arm** | | | **Comparison against treatment policy estimand (True HR = 1.00)** | |
| 2:1 allocation, 50% switching rate for control arm | Pure HE (100%) | 1.00 (0.85, 1.17) | -0.00 (-0.15, 0.17) | 0.86 |
| | Mixed TPE (25%) and HE (75%) | 1.00 (0.87, 1.15) | -0.00 (-0.13, 0.14) | 0.89 |
| | Mixed TPE (50%) and HE (50%) | 1.00 (0.88, 1.13) | -0.00 (-0.11, 0.12) | 0.91 |
| | Mixed TPE (75%) and HE (25%) | 1.00 (0.89, 1.12) | -0.00 (-0.10, 0.11) | 0.93 |
| | Pure TPE (100%) | 1.00 (0.90, 1.11) | -0.00 (-0.09, 0.10) | 0.96 |
| | Pure HE (100%) | 1.00 (0.86, 1.17) | 0.00 (-0.14, 0.16) | 0.87 |
| 1:1 allocation,<br>50% switching rate for control arm | Mixed TPE (25%) and HE (75%) | 1.00 (0.88, 1.14) | 0.00 (-0.12, 0.13) | 0.89 |
| | Mixed TPE (50%) and HE (50%) | 1.00 (0.89, 1.13) | 0.00 (-0.10, 0.12) | 0.92 |
| | Mixed TPE (75%) and HE (25%) | 1.00 (0.90, 1.12) | 0.00 (-0.09, 0.10) | 0.94 |
| | Pure TPE (100%) | 1.00 (0.91, 1.11) | 0.00 (-0.09, 0.10) | 0.96 |

**[a]This table shows estimated treatment effects under an assumed hazard ratio (HR) of 1.00 for the transition hazards of the illness-death model and bias and coverage in comparison to the true treatment policy estimand. Monte Carlo standard errors for all measures are very close to zero**
**Acronyms: CI: Confidence intervals; HE - Hypothetical estimator; HR - Hazard ratio; TPE - Treatment policy estimator.**

## Figures

*Figure 1: Distribution of HRs estimated under an assumed HR of 0.60 for the transition hazards of the illness-death model in the simulation with fixed-effects data-generating mechanism, random-effects meta-analysis, and rank-preserving structural failure time models. The dashed line indicates the true value of the treatment policy estimand.*

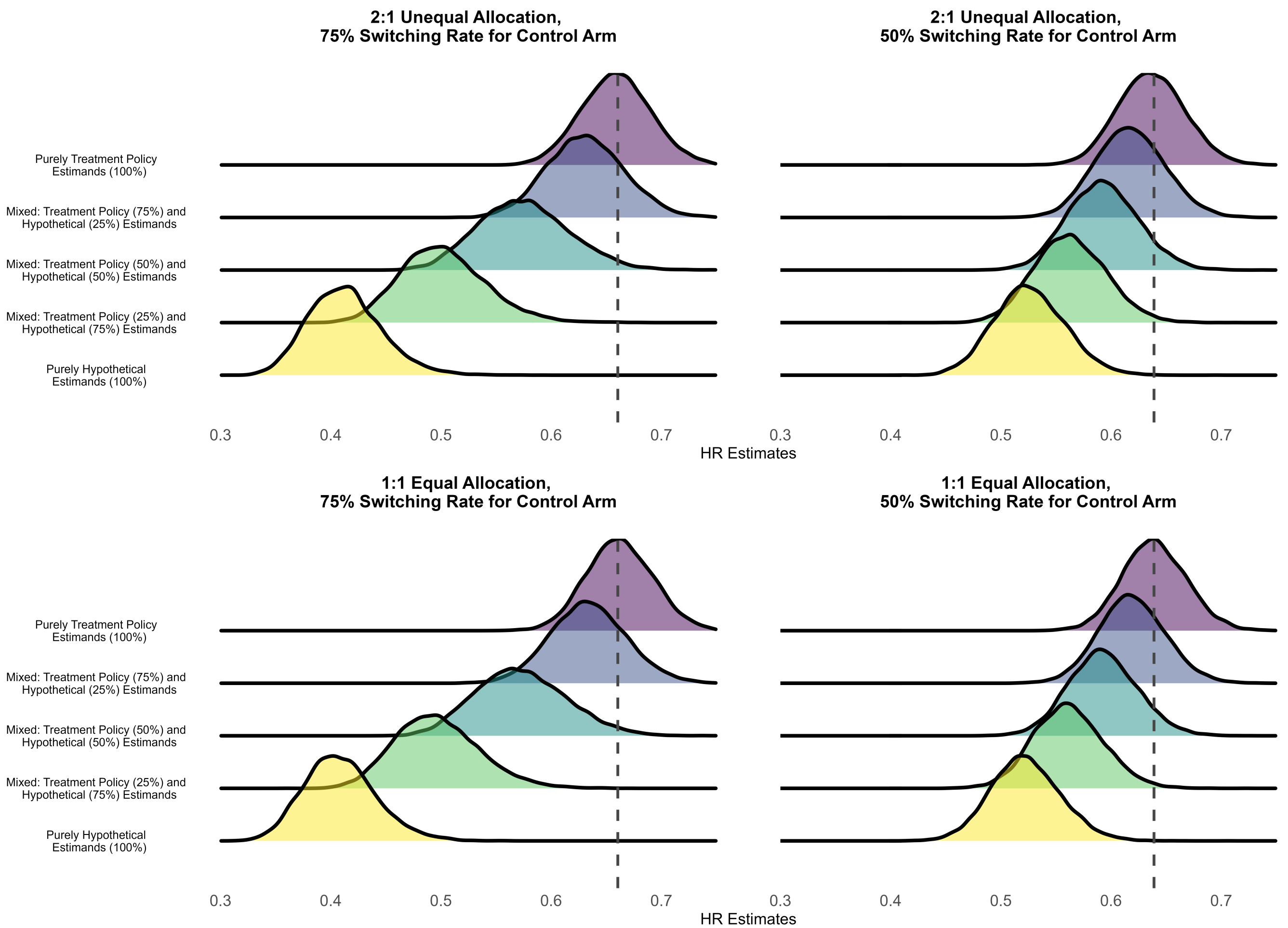

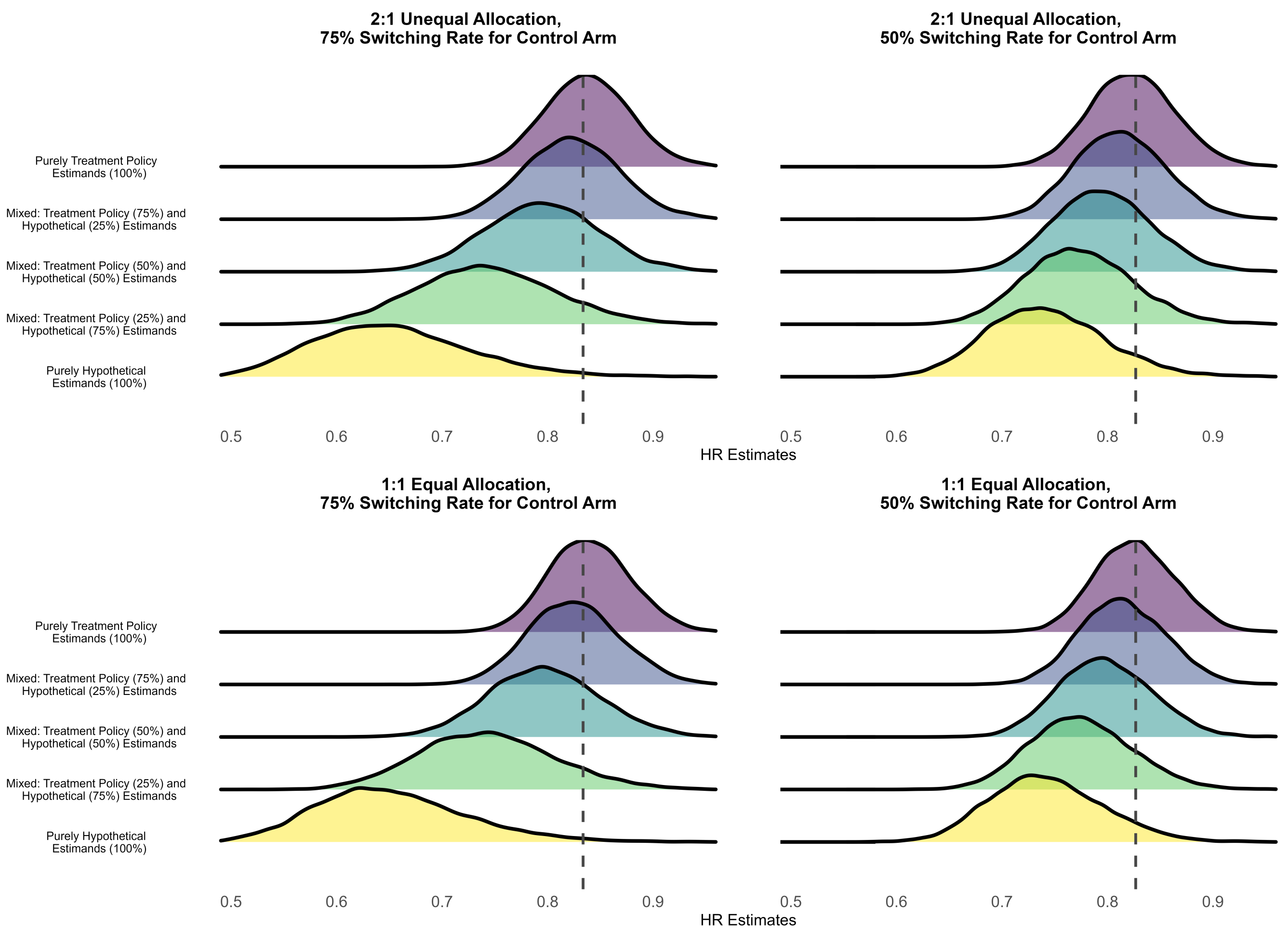

*Figure 2: Distribution of HRs estimated under an assumed HR of 0.80 for the transition hazards of the illness-death model in the simulation with fixed-effects data-generating mechanism, random-effects meta-analysis, and rank-preserving structural failure time models. The dashed line indicates the true value of the treatment policy estimand.*

*Figure 3: Distribution of HRs estimated under an assumed HR of 1.00 for the transition hazards of the illness-death model in the simulation with fixed-effects data-generating mechanism, random-effects meta-analysis, and rank-preserving structural failure time models. The dashed line indicates the true value of the treatment policy estimand.*

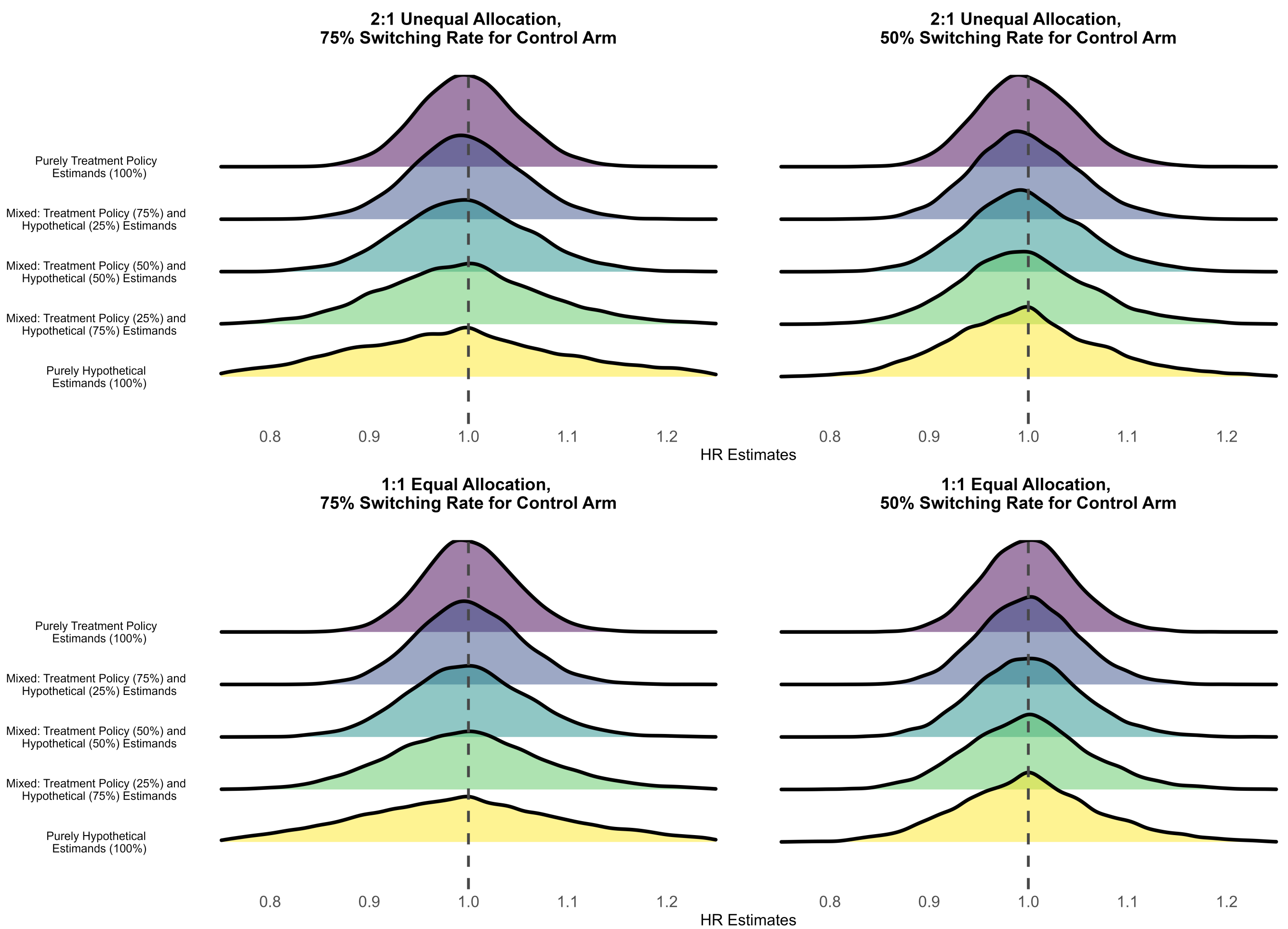

# Appendix of Estimands and Their Implications for Evidence Synthesis for Oncology: A Simulation Study of Treatment Switching in Meta-Analysis

**Overview**

In Appendix Section 1, the details of censoring switchers as an analytical strategy to estimate the hypothetical estimand are provided.

In Appendix Section 2, the results of the sensitivity analyses of the main simulation setting are presented. In particular, Appendix Section 2.1 presents the results where the hypothetical estimand is estimated using a Rank-Preserving Structural Failure Time Model (RPSFTM), under a fixed-effects data-generating mechanism and fixed-effects meta-analytical estimation. Appendix Section 2.2 presents the results where the hypothetical estimand is estimated using RPSFTM, under a random-effects data-generating mechanism and random-effects meta-analytical estimation.

In Appendix Section 3, the results of the supplementary simulations where switchers were censored at the time of switch to estimate the hypothetical estimand are presented. Appendix Section 3.1 presents the results under a fixed-effects data-generating mechanism and random-effects meta-analytical estimation. Appendix Section 3.2 presents results under a fixed-effects data-generating mechanism and fixed-effects meta-analytical estimation; and Appendix Section 3.3 presents results under a random-effects data-generating mechanism and random-effects meta-analytical estimation.

## 1. Details of censoring switchers

In the supplementary simulations, censoring switchers was used as the analytical strategy to estimate the hypothetical estimand at the trial level. In this strategy, a Cox proportional hazards model with OS as the outcome and treatment as a covariate was fit to the OS times of patients in the control and treatment groups, but the OS times of switchers in the control group were censored at the time of switching. We extracted the estimated log HR and standard error using the same approach as in the estimation of the treatment policy estimand.

## 2. Sensitivity analyses of main simulation

### *2.1 Fixed-effects data-generating mechanism, fixed-effects meta-analysis*

In this section, we present the results of the sensitivity analysis of the main simulation setting where estimates were pooled using a fixed-effects meta-analysis. A fixed-effects model was assumed when generating the trial data.

*Table 1: Averages of pooled treatment effect estimates and comparison against treatment policy estimand under an assumed HR of 0.60 for the transition hazards of the illness-death model in the simulation with fixed-effects data-generating mechanism, fixed-effects meta-analysis, and rank-preserving structural failure time models*

| **Scenarios** | **Estimators** | **Estimated treatment effects: HR (averaged 95% CI)** | **Bias (2.5%, 97.5% difference percentiles)** | **Coverage** |
|---|---|---|---|---|
| **75% switching rate for control arm** | | | **Comparison against treatment policy estimand (True HR = 0.66)** | |
| 2:1 allocation, 75% switching rate for control arm[a] | Pure HE (100%) | 0.41 (0.34, 0.51) | -0.25 (-0.30, -0.17) | 0.00 |
| | Mixed TPE (25%) and HE (75%) | 0.55 (0.48, 0.64) | -0.11 (-0.18, -0.02) | 0.31 |
| | Mixed TPE (50%) and HE (50%) | 0.61 (0.54, 0.69) | -0.05 (-0.12, 0.02) | 0.72 |
| | Mixed TPE (75%) and HE (25%) | 0.64 (0.58, 0.71) | -0.02 (-0.09, 0.05) | 0.90 |
| | Pure TPE (100%) | 0.66 (0.60, 0.72) | -0.00 (-0.06, 0.06) | 0.95 |
| 1:1 allocation, 75% switching rate for control arm | Pure HE (100%) | 0.41 (0.33, 0.50) | -0.25 (-0.31, -0.18) | 0.00 |
| | Mixed TPE (25%) and HE (75%) | 0.55 (0.48, 0.64) | -0.11 (-0.18, -0.03) | 0.29 |
| | Mixed TPE (50%) and HE (50%) | 0.61 (0.54, 0.69) | -0.05 (-0.12, 0.03) | 0.73 |
| | Mixed TPE (75%) and HE (25%) | 0.64 (0.58, 0.71) | -0.02 (-0.08, 0.05) | 0.92 |
| | Pure TPE (100%) | 0.66 (0.61, 0.72) | 0.00 (-0.05, 0.06) | 0.95 |
| **50% switching rate for control arm** | | | **Comparison against treatment policy estimand (True HR = 0.64)** | |
| | Pure HE (100%) | 0.53 (0.46, 0.60) | -0.11 (-0.17, -0.05) | 0.15 |
| 2:1 allocation, 50% switching rate for control arm | Mixed TPE (25%) and HE (75%) | 0.57 (0.50, 0.64) | -0.07 (-0.13, -0.00) | 0.49 |
| | Mixed TPE (50%) and HE (50%) | 0.60 (0.54, 0.67) | -0.04 (-0.10, 0.02) | 0.76 |
| | Mixed TPE (75%) and HE (25%) | 0.62 (0.56, 0.68) | -0.02 (-0.08, 0.04) | 0.90 |
| | Pure TPE (100%) | 0.64 (0.58, 0.70) | -0.00 (-0.06, 0.06) | 0.95 |
| 1:1 allocation, 50% switching rate for control arm | Pure HE (100%) | 0.52 (0.46, 0.59) | -0.12 (-0.18, -0.05) | 0.11 |
| | Mixed TPE (25%) and HE (75%) | 0.57 (0.51, 0.64) | -0.07 (-0.13, -0.01) | 0.45 |
| | Mixed TPE (50%) and HE (50%) | 0.60 (0.54, 0.66) | -0.04 (-0.10, 0.02) | 0.77 |
| | Mixed TPE (75%) and HE (25%) | 0.62 (0.57, 0.68) | -0.02 (-0.07, 0.04) | 0.91 |
| | Pure TPE (100%) | 0.64 (0.59, 0.70) | 0.00 (-0.06, 0.06) | 0.95 |

**[a]This table shows estimated treatment effects under an assumed hazard ratio (HR) of 0.60 for the transition hazards of the illness-death model and bias and coverage in comparison to the true treatment policy estimand. Monte Carlo standard errors for all measures are very close to zero**
**Acronyms: CI: Confidence intervals; HE - Hypothetical estimator; HR - Hazard ratio; TPE - Treatment policy estimator.**

*Table 2: Averages of pooled treatment effect estimates and comparison against treatment policy estimand under an assumed HR of 0.80 for the transition hazards of the illness-death model in the simulation with fixed-effects data-generating mechanism, fixed-effects meta-analysis, and rank-preserving structural failure time models*

| Scenarios | Estimators | Estimated treatment effects: HR (averaged 95% CI) | Bias (2.5%, 97.5% difference percentiles) | Coverage |
|---|---|---|---|---|
| **75% switching rate for control arm** | | | **Comparison against treatment policy estimand (True HR = 0.84)** | |
| 2:1 allocation, 75% switching rate for control arm[a] | Pure HE (100%) | 0.66 (0.53, 0.82) | -0.18 (-0.32, 0.00) | 0.36 |
| | Mixed TPE (25%) and HE (75%) | 0.77 (0.66, 0.89) | -0.07 (-0.19, 0.06) | 0.75 |
| | Mixed TPE (50%) and HE (50%) | 0.81 (0.71, 0.91) | -0.03 (-0.13, 0.07) | 0.87 |
| | Mixed TPE (75%) and HE (25%) | 0.83 (0.74, 0.92) | -0.01 (-0.10, 0.08) | 0.92 |
| | Pure TPE (100%) | 0.84 (0.76, 0.92) | -0.00 (-0.08, 0.08) | 0.95 |
| 1:1 allocation, 75% switching rate for control arm | Pure HE (100%) | 0.66 (0.53, 0.81) | -0.18 (-0.31, -0.00) | 0.34 |
| | Mixed TPE (25%) and HE (75%) | 0.77 (0.67, 0.89) | -0.07 (-0.18, 0.06) | 0.75 |
| | Mixed TPE (50%) and HE (50%) | 0.81 (0.72, 0.91) | -0.03 (-0.12, 0.07) | 0.89 |
| | Mixed TPE (75%) and HE (25%) | 0.83 (0.75, 0.92) | -0.01 (-0.09, 0.08) | 0.93 |
| | Pure TPE (100%) | 0.84 (0.77, 0.92) | 0.00 (-0.07, 0.08) | 0.95 |
| **50% switching rate for control arm** | | | **Comparison against treatment policy estimand (True HR = 0.83)** | |
| 2:1 allocation, 50% switching rate for control arm | Pure HE (100%) | 0.75 (0.65, 0.86) | -0.08 (-0.18, 0.07) | 0.63 |
| | Mixed TPE (25%) and HE (75%) | 0.78 (0.69, 0.88) | -0.05 (-0.14, 0.07) | 0.79 |
| | Mixed TPE (50%) and HE (50%) | 0.80 (0.72, 0.89) | -0.03 (-0.11, 0.08) | 0.87 |
| | Mixed TPE (75%) and HE (25%) | 0.81 (0.74, 0.90) | -0.01 (-0.09, 0.08) | 0.92 |
| | Pure TPE (100%) | 0.82 (0.75, 0.91) | -0.00 (-0.08, 0.08) | 0.95 |
| | Pure HE (100%) | 0.75 (0.65, 0.86) | -0.08 (-0.18, 0.05) | 0.63 |
| 1:1 allocation,<br>50% switching rate for control arm | Mixed TPE (25%) and HE (75%) | 0.78 (0.69, 0.88) | -0.04 (-0.14, 0.06) | 0.80 |
| | Mixed TPE (50%) and HE (50%) | 0.80 (0.72, 0.89) | -0.02 (-0.11, 0.07) | 0.89 |
| | Mixed TPE (75%) and HE (25%) | 0.82 (0.74, 0.90) | -0.01 (-0.09, 0.07) | 0.93 |
| | Pure TPE (100%) | 0.83 (0.76, 0.90) | 0.00 (-0.07, 0.08) | 0.95 |

**[a]This table shows estimated treatment effects under an assumed hazard ratio (HR) of 0.80 for the transition hazards of the illness-death model and bias and coverage in comparison to the true treatment policy estimand. Monte Carlo standard errors for all measures are very close to zero**
**Acronyms: CI: Confidence intervals; HE - Hypothetical estimator; HR - Hazard ratio; TPE - Treatment policy estimator.**

*Table 3: Averages of pooled treatment effect estimates and comparison against treatment policy estimand under an assumed HR of 1.00 for the transition hazards of the illness-death model in the simulation with fixed-effects data-generating mechanism, fixed-effects meta-analysis, and rank-preserving structural failure time models*

| **Scenarios** | **Estimators** | **Estimated treatment effects: HR (averaged 95% CI)** | **Bias (2.5%, 97.5% difference percentiles)** | **Coverage** |
|---|---|---|---|---|
| **75% switching rate for control arm** | | | **Comparison against treatment policy estimand (True HR = 1.00)** | |
| 2:1 allocation, 75% switching rate for control arm[a] | Pure HE (100%) | 0.99 (0.79, 1.24) | -0.01 (-0.23, 0.25) | 0.88 |
| | Mixed TPE (25%) and HE (75%) | 0.99 (0.85, 1.16) | -0.01 (-0.15, 0.16) | 0.90 |
| | Mixed TPE (50%) and HE (50%) | 1.00 (0.88, 1.13) | -0.00 (-0.12, 0.13) | 0.92 |
| | Mixed TPE (75%) and HE (25%) | 1.00 (0.90, 1.11) | -0.00 (-0.11, 0.11) | 0.93 |
| | Pure TPE (100%) | 1.00 (0.91, 1.10) | -0.00 (-0.10, 0.10) | 0.95 |
| 1:1 allocation, 75% switching rate for control arm | Pure HE (100%) | 1.00 (0.80, 1.24) | -0.00 (-0.21, 0.25) | 0.89 |
| | Mixed TPE (25%) and HE (75%) | 1.00 (0.86, 1.16) | -0.00 (-0.14, 0.16) | 0.90 |
| | Mixed TPE (50%) and HE (50%) | 1.00 (0.89, 1.13) | 0.00 (-0.11, 0.13) | 0.92 |
| | Mixed TPE (75%) and HE (25%) | 1.00 (0.90, 1.11) | 0.00 (-0.10, 0.11) | 0.93 |
| | Pure TPE (100%) | 1.00 (0.91, 1.10) | 0.00 (-0.09, 0.10) | 0.95 |
| **50% switching rate for control arm** | | | **Comparison against treatment policy estimand (True HR = 1.00)** | |
| 2:1 allocation, 50% switching rate for control arm | Pure HE (100%) | 0.99 (0.86, 1.15) | -0.01 (-0.15, 0.16) | 0.85 |
| | Mixed TPE (25%) and HE (75%) | 1.00 (0.88, 1.13) | -0.00 (-0.13, 0.14) | 0.88 |
| | Mixed TPE (50%) and HE (50%) | 1.00 (0.89, 1.12) | -0.00 (-0.11, 0.12) | 0.90 |
| | Mixed TPE (75%) and HE (25%) | 1.00 (0.90, 1.11) | -0.00 (-0.10, 0.11) | 0.92 |
| | Pure TPE (100%) | 1.00 (0.91, 1.10) | -0.00 (-0.09, 0.10) | 0.95 |
| | Pure HE (100%) | 1.00 (0.87, 1.15) | -0.00 (-0.14, 0.16) | 0.86 |
| 1:1 allocation,<br>50% switching rate for control arm | Mixed TPE (25%) and HE (75%) | 1.00 (0.89, 1.13) | 0.00 (-0.12, 0.13) | 0.88 |
| | Mixed TPE (50%) and HE (50%) | 1.00 (0.90, 1.11) | 0.00 (-0.10, 0.11) | 0.91 |
| | Mixed TPE (75%) and HE (25%) | 1.00 (0.91, 1.10) | 0.00 (-0.09, 0.10) | 0.93 |
| | Pure TPE (100%) | 1.00 (0.91, 1.10) | 0.00 (-0.09, 0.09) | 0.95 |

**[a]This table shows estimated treatment effects under an assumed hazard ratio (HR) of 1.00 for the transition hazards of the illness-death model and bias and coverage in comparison to the true treatment policy estimand. Monte Carlo standard errors for all measures are very close to zero**
**Acronyms: CI: Confidence intervals; HE - Hypothetical estimantor; HR - Hazard ratio; TPE - Treatment policy estimantor.**

*Figure 1: Distribution of HRs estimated under an assumed HR of 0.60 for the transition hazards of the illness-death model in the simulation with fixed-effects data-generating mechanism, fixed-effects meta-analysis, and rank-preserving structural failure time models. The dashed line indicates the true value of the treatment policy estimand.*

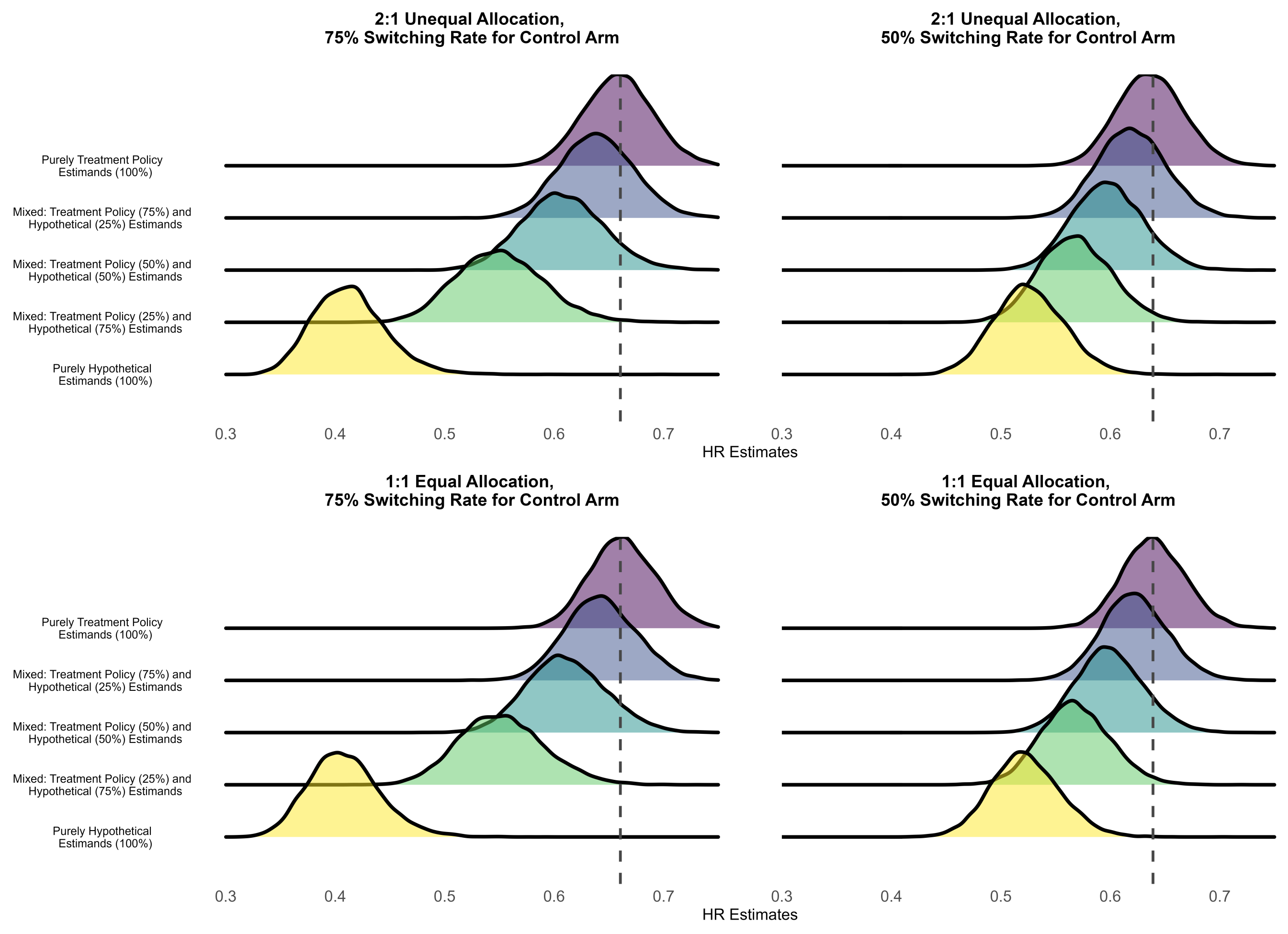

*Figure 2: Distribution of HRs estimated under an assumed HR of 0.80 for the transition hazards of the illness-death model in the simulation with fixed-effects data-generating mechanism, fixed-effects meta-analysis, and rank-preserving structural failure time models. The dashed line indicates the true value of the treatment policy estimand.*

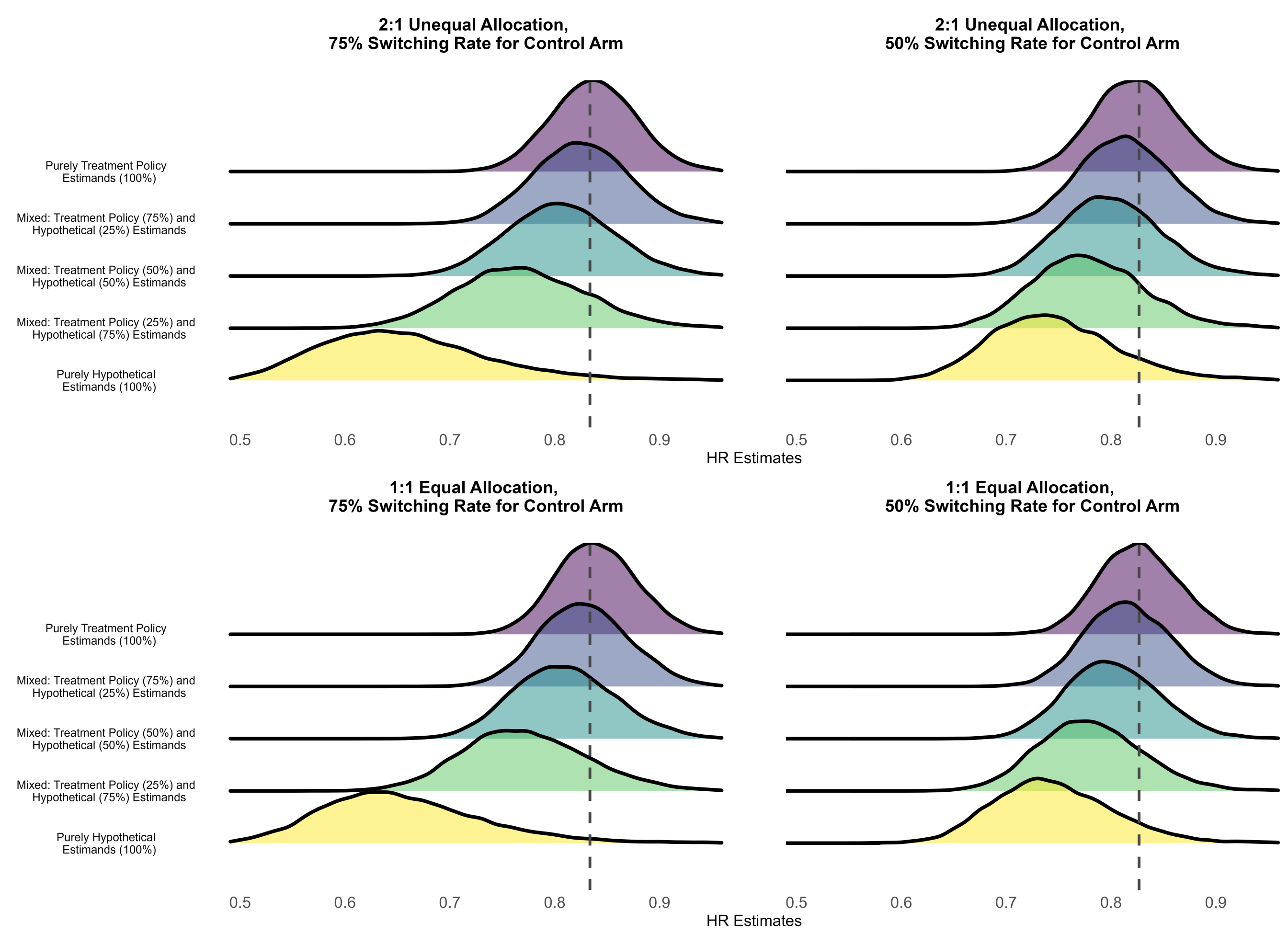

*Figure 3: Distribution of HRs estimated under an assumed HR of 1.00 for the transition hazards of the illness-death model in the simulation with fixed-effects data-generating mechanism, fixed-effects meta-analysis, and rank-preserving structural failure time models. The dashed line indicates the true value of the treatment policy estimand.*

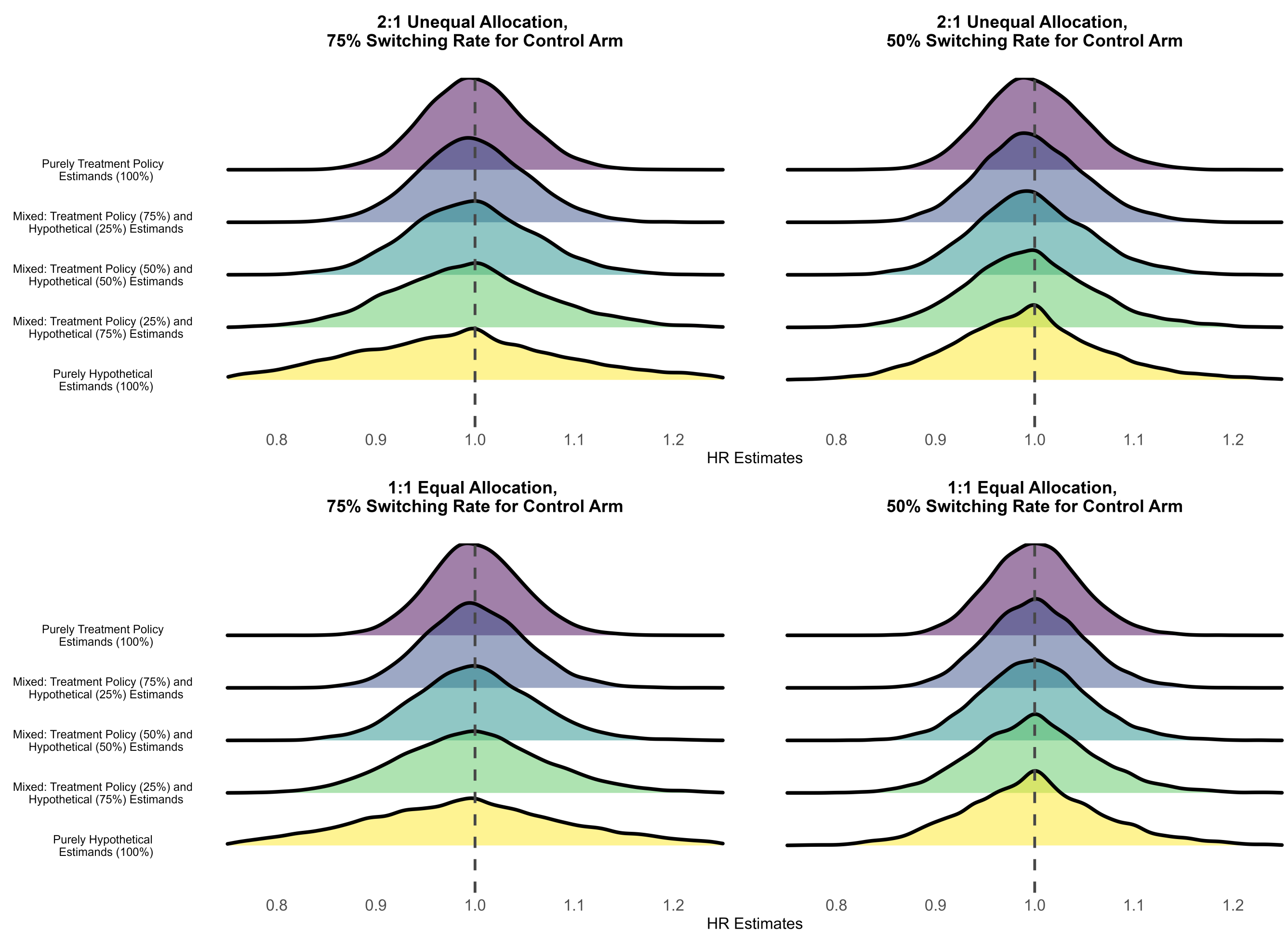

*2.2 Random-effects data-generating mechanism, random-effects meta-analysis*

In this section, we present the results of the sensitivity analysis of the main simulation setting, where trials were simulated assuming a random-effects model within each simulation replicate. Here, for $n$ trials in a replicate under a scenario where the transition HR was $\beta$, we sampled $n$ study-specific log treatment effects $\log\beta_1, \dots, \log\beta_n$ from $N(\log\beta, \tau^2)$ for a pre-selected $\tau^2$ of 0.03. Then the individual trials were simulated as above using each $\beta_i$ as the specified transition HR. The sample size in each trial was randomly chosen to be 250, 300, or 350 with equal probability. For each scenario, we repeated the simulation 10,000 times, corresponding to 80,000 simulated trials. The trial estimates were pooled assuming a random-effects model.

*Table 4: Averages of pooled treatment effect estimates and comparison against treatment policy estimand under an assumed HR of 0.60 for the transition hazards of the illness-death model in the simulation with random-effects data-generating mechanism, random-effects meta-analysis, and rank-preserving structural failure time models*

| **Scenarios** | **Estimators** | **Estimated treatment effects: HR (averaged 95% CI)** | **Bias (2.5%, 97.5% difference percentiles)** | **Coverage** |
|---|---|---|---|---|
| **75% switching rate for control arm** | | | **Comparison against treatment policy estimand (True HR = 0.66)** | |
| 2:1 allocation, 75% switching rate for control arm[a] | Pure HE (100%) | 0.44 (0.34, 0.57) | -0.22 (-0.31, -0.09) | 0.14 |
| | Mixed TPE (25%) and HE (75%) | 0.53 (0.41, 0.68) | -0.13 (-0.23, -0.01) | 0.54 |
| | Mixed TPE (50%) and HE (50%) | 0.59 (0.48, 0.72) | -0.07 (-0.17, 0.04) | 0.81 |
| | Mixed TPE (75%) and HE (25%) | 0.63 (0.54, 0.74) | -0.03 (-0.12, 0.08) | 0.90 |
| | Pure TPE (100%) | 0.66 (0.58, 0.75) | -0.00 (-0.09, 0.10) | 0.92 |
| 1:1 allocation, 75% switching rate for control arm | Pure HE (100%) | 0.44 (0.34, 0.56) | -0.22 (-0.31, -0.10) | 0.13 |
| | Mixed TPE (25%) and HE (75%) | 0.52 (0.40, 0.67) | -0.14 (-0.23, -0.01) | 0.52 |
| | Mixed TPE (50%) and HE (50%) | 0.58 (0.47, 0.72) | -0.08 (-0.17, 0.05) | 0.81 |
| | Mixed TPE (75%) and HE (25%) | 0.63 (0.54, 0.74) | -0.03 (-0.12, 0.08) | 0.91 |
| | Pure TPE (100%) | 0.66 (0.58, 0.75) | 0.00 (-0.08, 0.10) | 0.92 |
| **50% switching rate for control arm** | | | **Comparison against treatment policy estimand (True HR = 0.64)** | |
| 2:1 allocation,<br>50% switching rate for control arm | Pure HE (100%) | 0.53 (0.44, 0.63) | -0.11 (-0.20, -0.00) | 0.41 |
| | Mixed TPE (25%) and HE (75%) | 0.56 (0.47, 0.67) | -0.08 (-0.17, 0.03) | 0.65 |
| | Mixed TPE (50%) and HE (50%) | 0.59 (0.50, 0.70) | -0.05 (-0.14, 0.05) | 0.82 |
| | Mixed TPE (75%) and HE (25%) | 0.61 (0.53, 0.72) | -0.03 (-0.11, 0.08) | 0.89 |
| | Pure TPE (100%) | 0.64 (0.55, 0.73) | -0.00 (-0.09, 0.09) | 0.92 |
| 1:1 allocation,<br>50% switching rate for control arm | Pure HE (100%) | 0.53 (0.44, 0.63) | -0.11 (-0.20, -0.01) | 0.39 |
| | Mixed TPE (25%) and HE (75%) | 0.56 (0.47, 0.67) | -0.08 (-0.17, 0.02) | 0.65 |
| | Mixed TPE (50%) and HE (50%) | 0.59 (0.50, 0.70) | -0.05 (-0.14, 0.05) | 0.82 |
| | Mixed TPE (75%) and HE (25%) | 0.62 (0.53, 0.72) | -0.02 (-0.11, 0.08) | 0.90 |
| | Pure TPE (100%) | 0.64 (0.56, 0.73) | 0.00 (-0.09, 0.10) | 0.92 |

**[a]This table shows estimated treatment effects under an assumed hazard ratio (HR) of 0.60 for the transition hazards of the illness-death model and bias and coverage in comparison to the true treatment policy estimand. Monte Carlo standard errors for all measures are very close to zero**
**Acronyms: CI: Confidence intervals; HE - Hypothetical estimator; HR - Hazard ratio; TPE - Treatment policy estimator.**

*Table 5: Averages of pooled treatment effect estimates and comparison against treatment policy estimand under an assumed HR of 0.80 for the transition hazards of the illness-death model in the simulation with random-effects data-generating mechanism, random-effects meta-analysis, and rank-preserving structural failure time models*

| **Scenarios** | **Estimators** | **Estimated treatment effects: HR (averaged 95% CI)** | **Bias (2.5%, 97.5% difference percentiles)** | **Coverage** |
|---|---|---|---|---|
| **75% switching rate for control arm** | | | **Comparison against treatment policy estimand (True HR = 0.84)** | |
| 2:1 allocation, 75% switching rate for control arm[a] | Pure HE (100%) | 0.66 (0.49, 0.89) | -0.18 (-0.35, 0.06) | 0.55 |
| | Mixed TPE (25%) and HE (75%) | 0.73 (0.57, 0.93) | -0.11 (-0.27, 0.08) | 0.75 |
| | Mixed TPE (50%) and HE (50%) | 0.78 (0.64, 0.95) | -0.06 (-0.20, 0.10) | 0.86 |
| | Mixed TPE (75%) and HE (25%) | 0.81 (0.70, 0.95) | -0.03 (-0.15, 0.11) | 0.91 |
| | Pure TPE (100%) | 0.84 (0.73, 0.96) | -0.00 (-0.11, 0.12) | 0.92 |
| 1:1 allocation, 75% switching rate for control arm | Pure HE (100%) | 0.66 (0.49, 0.89) | -0.18 (-0.35, 0.05) | 0.56 |
| | Mixed TPE (25%) and HE (75%) | 0.73 (0.57, 0.93) | -0.11 (-0.27, 0.08) | 0.76 |
| | Mixed TPE (50%) and HE (50%) | 0.78 (0.64, 0.95) | -0.06 (-0.20, 0.10) | 0.87 |
| | Mixed TPE (75%) and HE (25%) | 0.82 (0.70, 0.95) | -0.02 (-0.14, 0.11) | 0.91 |
| | Pure TPE (100%) | 0.84 (0.74, 0.95) | -0.00 (-0.10, 0.11) | 0.92 |
| **50% switching rate for control arm** | | | **Comparison against treatment policy estimand (True HR = 0.83)** | |
| 2:1 allocation, 50% switching rate for control arm | Pure HE (100%) | 0.74 (0.61, 0.91) | -0.08 (-0.23, 0.09) | 0.73 |
| | Mixed TPE (25%) and HE (75%) | 0.77 (0.64, 0.93) | -0.06 (-0.19, 0.10) | 0.81 |
| | Mixed TPE (50%) and HE (50%) | 0.79 (0.67, 0.94) | -0.04 (-0.16, 0.11) | 0.87 |
| | Mixed TPE (75%) and HE (25%) | 0.81 (0.69, 0.94) | -0.02 (-0.14, 0.12) | 0.90 |
| | Pure TPE (100%) | 0.82 (0.72, 0.95) | -0.00 (-0.11, 0.12) | 0.92 |
| | Pure HE (100%) | 0.74 (0.61, 0.91) | -0.08 (-0.22, 0.09) | 0.72 |
| 1:1 allocation,<br>50% switching rate for control arm | Mixed TPE (25%) and HE (75%) | 0.77 (0.64, 0.93) | -0.06 (-0.19, 0.10) | 0.81 |
| | Mixed TPE (50%) and HE (50%) | 0.79 (0.67, 0.94) | -0.03 (-0.16, 0.11) | 0.87 |
| | Mixed TPE (75%) and HE (25%) | 0.81 (0.70, 0.94) | -0.02 (-0.13, 0.11) | 0.90 |
| | Pure TPE (100%) | 0.82 (0.72, 0.95) | -0.00 (-0.11, 0.12) | 0.92 |

**[a]This table shows estimated treatment effects under an assumed hazard ratio (HR) of 0.80 for the transition hazards of the illness-death model and bias and coverage in comparison to the true treatment policy estimand. Monte Carlo standard errors for all measures are very close to zero**
**Acronyms: CI: Confidence intervals; HE - Hypothetical estimator; HR - Hazard ratio; TPE - Treatment policy estimator.**

*Table 6: Averages of pooled treatment effect estimates and comparison against treatment policy estimand under an assumed HR of 1.00 for the transition hazards of the illness-death model in the simulation with random-effects data-generating mechanism, random-effects meta-analysis, and rank-preserving structural failure time models*

| **Scenarios** | **Estimators** | **Estimated treatment effects: HR (averaged 95% CI)** | **Bias (2.5%, 97.5% difference percentiles)** | **Coverage** |
|---|---|---|---|---|
| **75% switching rate for control arm** | | | **Comparison against treatment policy estimand (True HR = 1.00)** | |
| 2:1 allocation, 75% switching rate for control arm[a] | Pure HE (100%) | 0.98 (0.72, 1.35) | -0.02 (-0.30, 0.34) | 0.87 |
| | Mixed TPE (25%) and HE (75%) | 0.98 (0.78, 1.25) | -0.02 (-0.23, 0.24) | 0.88 |
| | Mixed TPE (50%) and HE (50%) | 0.99 (0.82, 1.19) | -0.01 (-0.18, 0.18) | 0.90 |
| | Mixed TPE (75%) and HE (25%) | 0.99 (0.85, 1.16) | -0.01 (-0.15, 0.15) | 0.91 |
| | Pure TPE (100%) | 1.00 (0.87, 1.14) | -0.00 (-0.13, 0.13) | 0.92 |
| 1:1 allocation, 75% switching rate for control arm | Pure HE (100%) | 0.99 (0.73, 1.35) | -0.01 (-0.29, 0.35) | 0.87 |
| | Mixed TPE (25%) and HE (75%) | 0.99 (0.79, 1.25) | -0.01 (-0.22, 0.25) | 0.89 |
| | Mixed TPE (50%) and HE (50%) | 0.99 (0.83, 1.19) | -0.01 (-0.18, 0.19) | 0.90 |
| | Mixed TPE (75%) and HE (25%) | 1.00 (0.86, 1.16) | -0.00 (-0.14, 0.16) | 0.91 |
| | Pure TPE (100%) | 1.00 (0.88, 1.14) | -0.00 (-0.12, 0.13) | 0.92 |
| **50% switching rate for control arm** | | | **Comparison against treatment policy estimand (True HR = 1.00)** | |
| 2:1 allocation, 50% switching rate for control arm | Pure HE (100%) | 1.00 (0.80, 1.24) | -0.00 (-0.21, 0.25) | 0.85 |
| | Mixed TPE (25%) and HE (75%) | 1.00 (0.82, 1.21) | -0.00 (-0.18, 0.21) | 0.87 |
| | Mixed TPE (50%) and HE (50%) | 1.00 (0.84, 1.18) | -0.00 (-0.17, 0.18) | 0.89 |
| | Mixed TPE (75%) and HE (25%) | 1.00 (0.86, 1.16) | -0.00 (-0.15, 0.16) | 0.90 |
| | Pure TPE (100%) | 1.00 (0.87, 1.15) | -0.00 (-0.14, 0.14) | 0.92 |
| | Pure HE (100%) | 1.00 (0.81, 1.25) | 0.00 (-0.20, 0.25) | 0.86 |
| 1:1 allocation,<br>50% switching rate for control arm | Mixed TPE (25%) and HE (75%) | 1.00 (0.83, 1.21) | 0.00 (-0.18, 0.21) | 0.87 |
| | Mixed TPE (50%) and HE (50%) | 1.00 (0.85, 1.18) | 0.00 (-0.16, 0.18) | 0.89 |
| | Mixed TPE (75%) and HE (25%) | 1.00 (0.86, 1.16) | -0.00 (-0.14, 0.16) | 0.90 |
| | Pure TPE (100%) | 1.00 (0.87, 1.14) | -0.00 (-0.13, 0.14) | 0.92 |

**[a]This table shows estimated treatment effects under an assumed hazard ratio (HR) of 1.00 for the transition hazards of the illness-death model and bias and coverage in comparison to the true treatment policy estimand. Monte Carlo standard errors for all measures are very close to zero**
**Acronyms: CI: Confidence intervals; HE - Hypothetical estimator; HR - Hazard ratio; TPE - Treatment policy estimator.**

*Figure 4: Distribution of HRs estimated under an assumed HR of 0.60 for the transition hazards of the illness-death model in the simulation with random-effects data-generating mechanism, random-effects meta-analysis, and rank-preserving structural failure time models. The dashed line indicates the true value of the treatment policy estimand.*

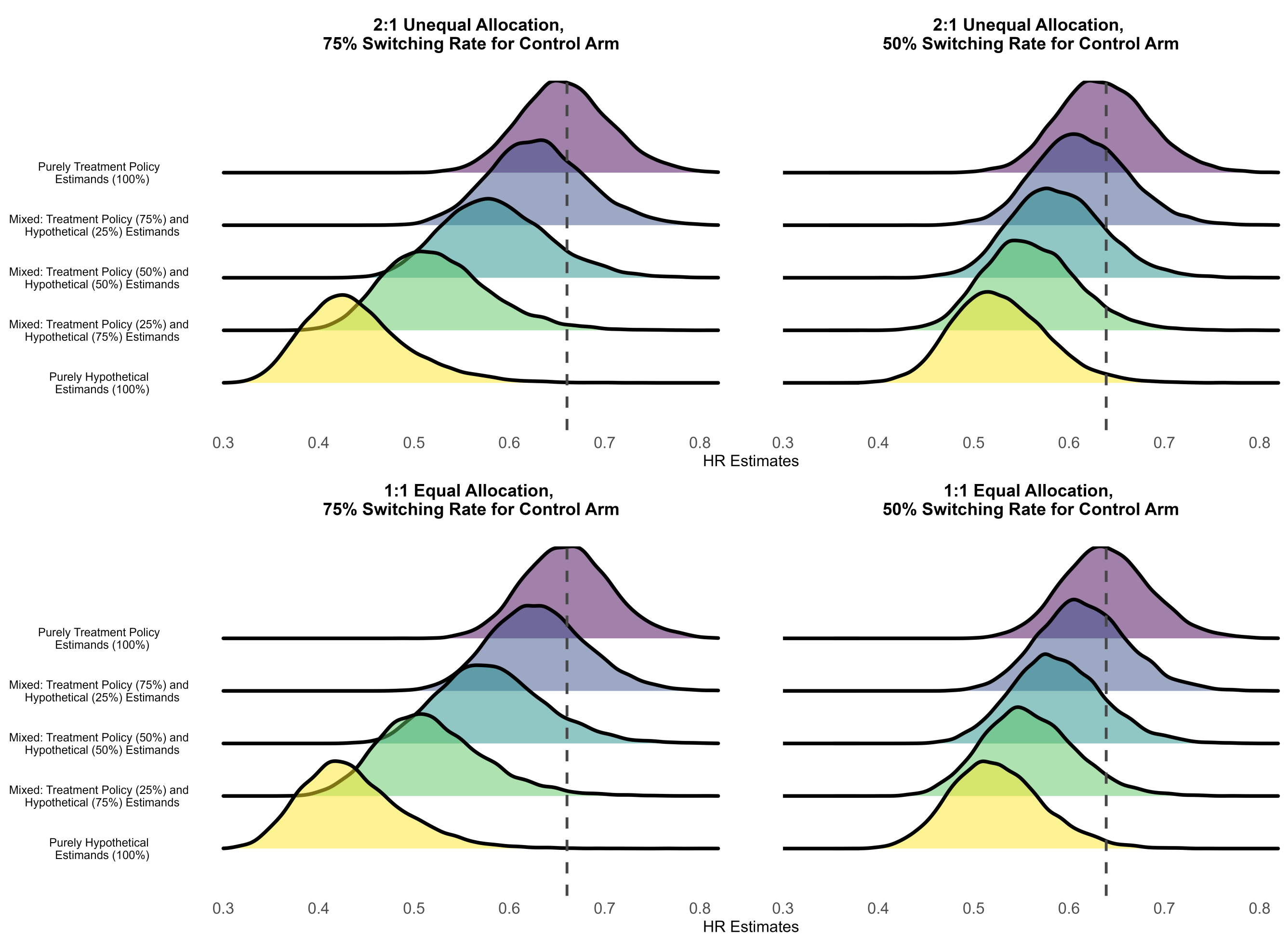

*Figure 5: Distribution of HRs estimated under an assumed HR of 0.80 for the transition hazards of the illness-death model in the simulation with random-effects data-generating mechanism, random-effects meta-analysis, and rank-preserving structural failure time models. The dashed line indicates the true value of the treatment policy estimand.*

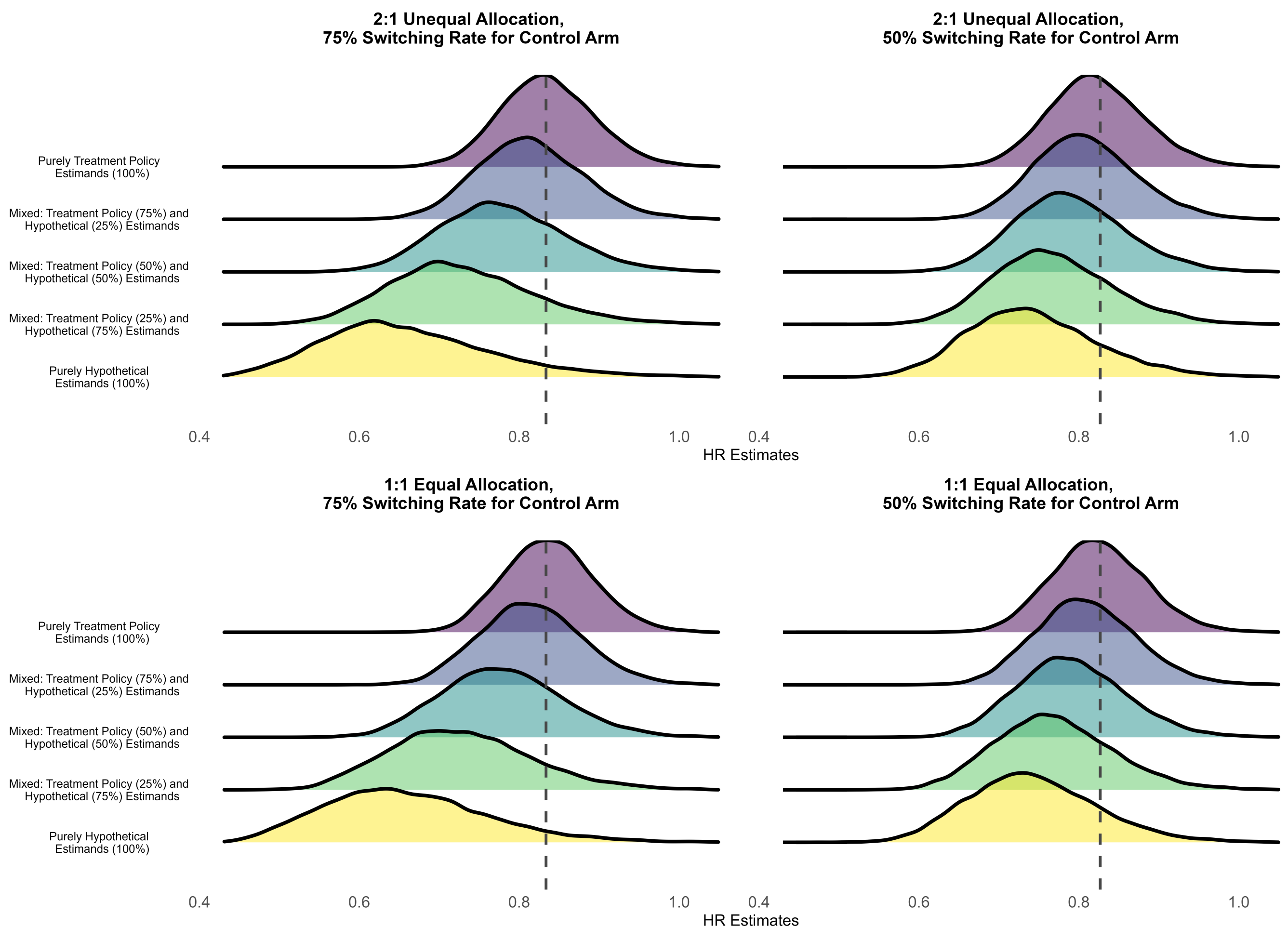

*Figure 6: Distribution of HRs estimated under an assumed HR of 1.00 for the transition hazards of the illness-death model in the simulation with random-effects data-generating mechanism, random-effects meta-analysis, and rank-preserving structural failure time models. The dashed line indicates the true value of the treatment policy estimand.*

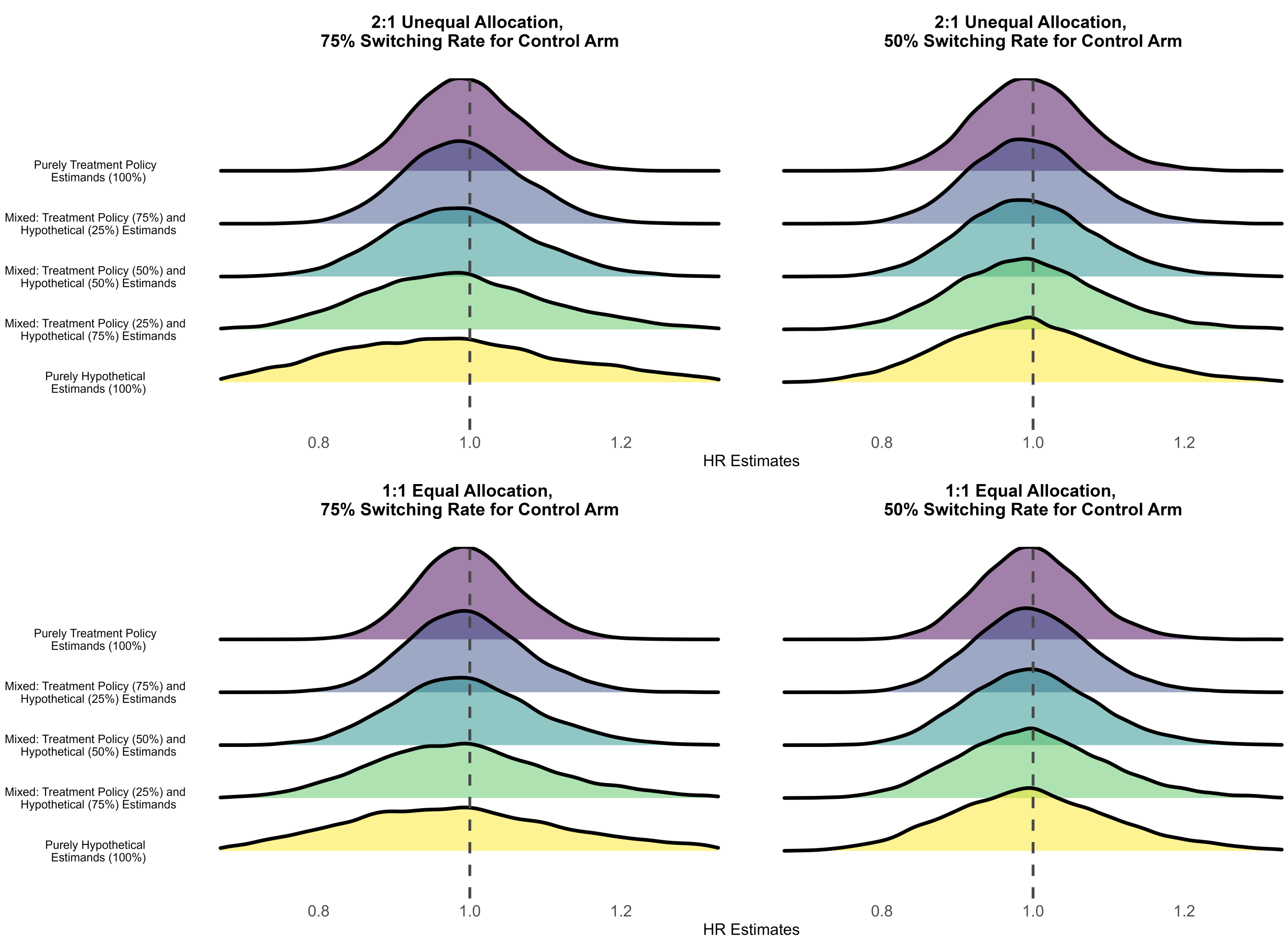

## 3. Supplementary simulations

In this section, we present the results of the supplementary simulation setting, where the analytical strategy to target the hypothetical estimand was censoring switchers instead of rank-preserving structural failure time models. The data-generating mechanism and meta-analysis model of each supplementary simulation setting were identical to the corresponding main simulation.

### *3.1 Fixed-effects data-generating mechanism, random-effects meta-analysis*

*Table 7: Averages of pooled treatment effect estimates and comparison against treatment policy estimand under an assumed HR of 0.60 for the transition hazards of the illness-death model in the simulation with fixed-effects data-generating mechanism, random-effects meta-analysis, and censoring switchers*

| Scenarios | Estimators | Estimated treatment effects: HR (averaged 95% CI) | Bias (2.5%, 97.5% difference percentiles) | Coverage |
|---|---|---|---|---|
| **75% switching rate for control arm** | | | **Comparison against treatment policy estimand (True HR = 0.66)** | |
| 2:1 allocation, 75% switching rate for control arm[a] | Pure HE (100%) | 0.61 (0.51, 0.72) | -0.05 (-0.14, 0.05) | 0.85 |
| | Mixed TPE (25%) and HE (75%) | 0.63 (0.54, 0.73) | -0.03 (-0.11, 0.05) | 0.91 |
| | Mixed TPE (50%) and HE (50%) | 0.64 (0.57, 0.73) | -0.02 (-0.09, 0.06) | 0.95 |
| | Mixed TPE (75%) and HE (25%) | 0.65 (0.58, 0.73) | -0.01 (-0.07, 0.06) | 0.96 |
| | Pure TPE (100%) | 0.66 (0.60, 0.73) | 0.00 (-0.06, 0.06) | 0.96 |
| 1:1 allocation, 75% switching rate for control arm | Pure HE (100%) | 0.61 (0.52, 0.72) | -0.05 (-0.13, 0.04) | 0.85 |
| | Mixed TPE (25%) and HE (75%) | 0.63 (0.55, 0.72) | -0.03 (-0.10, 0.05) | 0.92 |
| | Mixed TPE (50%) and HE (50%) | 0.65 (0.57, 0.73) | -0.01 (-0.08, 0.06) | 0.95 |
| | Mixed TPE (75%) and HE (25%) | 0.65 (0.59, 0.73) | -0.00 (-0.07, 0.06) | 0.96 |
| | Pure TPE (100%) | 0.66 (0.60, 0.73) | 0.00 (-0.05, 0.06) | 0.96 |
| **50% switching rate for control arm** | | | **Comparison against treatment policy estimand (True HR = 0.64)** | |
| | Pure HE (100%) | 0.60 (0.53, 0.69) | -0.04 (-0.11, 0.04) | 0.86 |
| 2:1 allocation, 50% switching rate for control arm | Mixed TPE (25%) and HE (75%) | 0.61 (0.54, 0.69) | -0.03 (-0.09, 0.04) | 0.91 |
| | Mixed TPE (50%) and HE (50%) | 0.62 (0.55, 0.70) | -0.02 (-0.08, 0.05) | 0.94 |
| | Mixed TPE (75%) and HE (25%) | 0.63 (0.57, 0.70) | -0.01 (-0.07, 0.06) | 0.95 |
| | Pure TPE (100%) | 0.64 (0.58, 0.71) | -0.00 (-0.06, 0.06) | 0.96 |
| 1:1 allocation, 50% switching rate for control arm | Pure HE (100%) | 0.60 (0.53, 0.68) | -0.04 (-0.10, 0.03) | 0.86 |
| | Mixed TPE (25%) and HE (75%) | 0.61 (0.55, 0.69) | -0.02 (-0.09, 0.04) | 0.91 |
| | Mixed TPE (50%) and HE (50%) | 0.62 (0.56, 0.70) | -0.01 (-0.07, 0.05) | 0.94 |
| | Mixed TPE (75%) and HE (25%) | 0.63 (0.57, 0.70) | -0.01 (-0.06, 0.06) | 0.96 |
| | Pure TPE (100%) | 0.64 (0.58, 0.71) | 0.00 (-0.06, 0.06) | 0.96 |

**[a]This table shows estimated treatment effects under an assumed hazard ratio (HR) of 0.60 for the transition hazards of the illness-death model and bias and coverage in comparison to the true treatment policy estimand. Monte Carlo standard errors for all measures are very close to zero**
**Acronyms: CI: Confidence intervals; HE - Hypothetical estimator; HR - Hazard ratio; TPE - Treatment policy estimator.**

*Table 8: Averages of pooled treatment effect estimates and comparison against treatment policy estimand under an assumed HR of 0.80 for the transition hazards of the illness-death model in the simulation with fixed-effects data-generating mechanism, random-effects meta-analysis, and censoring switchers*

| Scenarios | Estimators | Estimated treatment effects: HR (averaged 95% CI) | Bias (2.5%, 97.5% difference percentiles) | Coverage |
|---|---|---|---|---|
| **75% switching rate for control arm** | | | **Comparison against treatment policy estimand (True HR = 0.84)** | |
| 2:1 allocation, 75% switching rate for control arm[a] | Pure HE (100%) | 0.81 (0.68, 0.97) | -0.03 (-0.15, 0.10) | 0.94 |
| | Mixed TPE (25%) and HE (75%) | 0.82 (0.71, 0.95) | -0.02 (-0.12, 0.10) | 0.95 |
| | Mixed TPE (50%) and HE (50%) | 0.83 (0.73, 0.94) | -0.01 (-0.10, 0.09) | 0.96 |
| | Mixed TPE (75%) and HE (25%) | 0.83 (0.74, 0.93) | -0.01 (-0.09, 0.08) | 0.96 |
| | Pure TPE (100%) | 0.84 (0.76, 0.93) | -0.00 (-0.08, 0.08) | 0.96 |
| 1:1 allocation, 75% switching rate for control arm | Pure HE (100%) | 0.81 (0.69, 0.96) | -0.03 (-0.14, 0.10) | 0.94 |
| | Mixed TPE (25%) and HE (75%) | 0.82 (0.72, 0.95) | -0.02 (-0.12, 0.09) | 0.95 |
| | Mixed TPE (50%) and HE (50%) | 0.83 (0.74, 0.94) | -0.01 (-0.10, 0.09) | 0.96 |
| | Mixed TPE (75%) and HE (25%) | 0.84 (0.75, 0.93) | -0.00 (-0.08, 0.08) | 0.96 |
| | Pure TPE (100%) | 0.84 (0.76, 0.93) | 0.00 (-0.07, 0.08) | 0.96 |
| **50% switching rate for control arm** | | | **Comparison against treatment policy estimand (True HR = 0.83)** | |
| 2:1 allocation, 50% switching rate for control arm | Pure HE (100%) | 0.80 (0.70, 0.92) | -0.02 (-0.12, 0.08) | 0.93 |
| | Mixed TPE (25%) and HE (75%) | 0.81 (0.71, 0.92) | -0.02 (-0.11, 0.08) | 0.95 |
| | Mixed TPE (50%) and HE (50%) | 0.81 (0.73, 0.92) | -0.01 (-0.10, 0.08) | 0.95 |
| | Mixed TPE (75%) and HE (25%) | 0.82 (0.73, 0.91) | -0.01 (-0.09, 0.08) | 0.96 |
| | Pure TPE (100%) | 0.82 (0.74, 0.91) | -0.00 (-0.08, 0.08) | 0.96 |
| | Pure HE (100%) | 0.80 (0.71, 0.91) | -0.02 (-0.11, 0.07) | 0.94 |
| 1:1 allocation,<br>50% switching rate for control arm | Mixed TPE (25%) and HE (75%) | 0.81 (0.72, 0.91) | -0.02 (-0.10, 0.08) | 0.95 |
| | Mixed TPE (50%) and HE (50%) | 0.82 (0.73, 0.91) | -0.01 (-0.09, 0.08) | 0.96 |
| | Mixed TPE (75%) and HE (25%) | 0.82 (0.74, 0.91) | -0.00 (-0.08, 0.08) | 0.96 |
| | Pure TPE (100%) | 0.83 (0.75, 0.91) | 0.00 (-0.07, 0.08) | 0.96 |

**[a]This table shows estimated treatment effects under an assumed hazard ratio (HR) of 0.80 for the transition hazards of the illness-death model and bias and coverage in comparison to the true treatment policy estimand. Monte Carlo standard errors for all measures are very close to zero**
**Acronyms: CI: Confidence intervals; HE - Hypothetical estimator; HR - Hazard ratio; TPE - Treatment policy estimator.**

*Table 9: Averages of pooled treatment effect estimates and comparison against treatment policy estimand under an assumed HR of 1.00 for the transition hazards of the illness-death model in the simulation with fixed-effects data-generating mechanism, random-effects meta-analysis, and censoring switchers*

| Scenarios | Estimators | Estimated treatment effects: HR (averaged 95% CI) | Bias (2.5%, 97.5% difference percentiles) | Coverage |
|---|---|---|---|---|
| **75% switching rate for control arm** | | | **Comparison against treatment policy estimand (True HR = 1.00)** | |
| 2:1 allocation, 75% switching rate for control arm[a] | Pure HE (100%) | 1.01 (0.84, 1.21) | 0.01 (-0.14, 0.19) | 0.97 |
| | Mixed TPE (25%) and HE (75%) | 1.00 (0.87, 1.17) | 0.00 (-0.13, 0.15) | 0.96 |
| | Mixed TPE (50%) and HE (50%) | 1.00 (0.88, 1.14) | 0.00 (-0.11, 0.13) | 0.96 |
| | Mixed TPE (75%) and HE (25%) | 1.00 (0.89, 1.12) | 0.00 (-0.10, 0.11) | 0.96 |
| | Pure TPE (100%) | 1.00 (0.90, 1.11) | -0.00 (-0.09, 0.10) | 0.96 |
| 1:1 allocation, 75% switching rate for control arm | Pure HE (100%) | 1.02 (0.86, 1.20) | 0.02 (-0.13, 0.18) | 0.96 |
| | Mixed TPE (25%) and HE (75%) | 1.01 (0.88, 1.16) | 0.01 (-0.11, 0.15) | 0.96 |
| | Mixed TPE (50%) and HE (50%) | 1.01 (0.89, 1.14) | 0.01 (-0.10, 0.12) | 0.96 |
| | Mixed TPE (75%) and HE (25%) | 1.00 (0.90, 1.12) | 0.00 (-0.09, 0.11) | 0.96 |
| | Pure TPE (100%) | 1.00 (0.91, 1.11) | 0.00 (-0.09, 0.10) | 0.96 |
| **50% switching rate for control arm** | | | **Comparison against treatment policy estimand (True HR = 1.00)** | |
| 2:1 allocation, 50% switching rate for control arm | Pure HE (100%) | 1.00 (0.87, 1.15) | 0.00 (-0.12, 0.13) | 0.96 |
| | Mixed TPE (25%) and HE (75%) | 1.00 (0.88, 1.14) | 0.00 (-0.11, 0.12) | 0.96 |
| | Mixed TPE (50%) and HE (50%) | 1.00 (0.89, 1.13) | -0.00 (-0.11, 0.11) | 0.96 |
| | Mixed TPE (75%) and HE (25%) | 1.00 (0.89, 1.12) | -0.00 (-0.10, 0.11) | 0.96 |
| | Pure TPE (100%) | 1.00 (0.90, 1.11) | -0.00 (-0.09, 0.10) | 0.96 |
| | Pure HE (100%) | 1.00 (0.88, 1.14) | 0.00 (-0.11, 0.12) | 0.96 |
| 1:1 allocation,<br>50% switching rate for control arm | Mixed TPE (25%) and HE (75%) | 1.00 (0.89, 1.13) | 0.00 (-0.10, 0.11) | 0.97 |
| | Mixed TPE (50%) and HE (50%) | 1.00 (0.90, 1.12) | 0.00 (-0.10, 0.11) | 0.96 |
| | Mixed TPE (75%) and HE (25%) | 1.00 (0.90, 1.11) | 0.00 (-0.09, 0.10) | 0.96 |
| | Pure TPE (100%) | 1.00 (0.91, 1.11) | 0.00 (-0.09, 0.10) | 0.96 |

**[a]This table shows estimated treatment effects under an assumed hazard ratio (HR) of 1.00 for the transition hazards of the illness-death model and bias and coverage in comparison to the true treatment policy estimand. Monte Carlo standard errors for all measures are very close to zero**
**Acronyms: CI: Confidence intervals; HE - Hypothetical estimator; HR - Hazard ratio; TPE - Treatment policy estimator.**

### *3.2 Fixed-effects data-generating mechanism, fixed-effects meta-analysis*

*Table 10: Averages of pooled treatment effect estimates and comparison against treatment policy estimand under an assumed HR of 0.60 for the transition hazards of the illness-death model in the simulation with fixed-effects data-generating mechanism, fixed-effects meta-analysis, and censoring switchers*

| Scenarios | Estimators | Estimated treatment effects: HR (averaged 95% CI) | Bias (2.5%, 97.5% difference percentiles) | Coverage |
|---|---|---|---|---|
| **75% switching rate for control arm** | | | **Comparison against treatment policy estimand (True HR = 0.66)** | |
| 2:1 allocation, 75% switching rate for control arm[a] | Pure HE (100%) | 0.61 (0.52, 0.71) | -0.05 (-0.14, 0.04) | 0.81 |
| | Mixed TPE (25%) and HE (75%) | 0.63 (0.56, 0.72) | -0.03 (-0.10, 0.06) | 0.89 |
| | Mixed TPE (50%) and HE (50%) | 0.65 (0.58, 0.72) | -0.01 (-0.08, 0.06) | 0.93 |
| | Mixed TPE (75%) and HE (25%) | 0.65 (0.59, 0.72) | -0.01 (-0.07, 0.06) | 0.94 |
| | Pure TPE (100%) | 0.66 (0.60, 0.72) | 0.00 (-0.06, 0.06) | 0.95 |
| 1:1 allocation, 75% switching rate for control arm | Pure HE (100%) | 0.61 (0.53, 0.71) | -0.05 (-0.13, 0.05) | 0.82 |
| | Mixed TPE (25%) and HE (75%) | 0.63 (0.56, 0.72) | -0.02 (-0.10, 0.06) | 0.90 |
| | Mixed TPE (50%) and HE (50%) | 0.65 (0.58, 0.72) | -0.01 (-0.08, 0.06) | 0.94 |
| | Mixed TPE (75%) and HE (25%) | 0.66 (0.60, 0.72) | -0.00 (-0.06, 0.06) | 0.95 |
| | Pure TPE (100%) | 0.66 (0.61, 0.72) | 0.00 (-0.05, 0.06) | 0.95 |
| **50% switching rate for control arm** | | | **Comparison against treatment policy estimand (True HR = 0.64)** | |
| 2:1 allocation, 50% switching rate for control arm | Pure HE (100%) | 0.60 (0.53, 0.68) | -0.04 (-0.11, 0.04) | 0.82 |
| | Mixed TPE (25%) and HE (75%) | 0.61 (0.55, 0.69) | -0.03 (-0.09, 0.05) | 0.89 |
| | Mixed TPE (50%) and HE (50%) | 0.62 (0.56, 0.69) | -0.02 (-0.08, 0.05) | 0.92 |
| | Mixed TPE (75%) and HE (25%) | 0.63 (0.57, 0.70) | -0.01 (-0.07, 0.06) | 0.94 |
| | Pure TPE (100%) | 0.64 (0.58, 0.70) | -0.00 (-0.06, 0.06) | 0.95 |
| 1:1 allocation, 50% switching rate for control arm | Pure HE (100%) | 0.60 (0.54, 0.68) | -0.04 (-0.10, 0.03) | 0.83 |
| | Mixed TPE (25%) and HE (75%) | 0.62 (0.55, 0.68) | -0.02 (-0.09, 0.04) | 0.89 |
| | Mixed TPE (50%) and HE (50%) | 0.63 (0.57, 0.69) | -0.01 (-0.07, 0.05) | 0.92 |
| | Mixed TPE (75%) and HE (25%) | 0.63 (0.58, 0.70) | -0.01 (-0.06, 0.06) | 0.94 |
| | Pure TPE (100%) | 0.64 (0.59, 0.70) | 0.00 (-0.05, 0.06) | 0.95 |

**[a]This table shows estimated treatment effects under an assumed hazard ratio (HR) of 0.60 for the transition hazards of the illness-death model and bias and coverage in comparison to the true treatment policy estimand. Monte Carlo standard errors for all measures are very close to zero**
**Acronyms: CI: Confidence intervals; HE - Hypothetical estimator; HR - Hazard ratio; TPE - Treatment policy estimator.**

*Table 11: Averages of pooled treatment effect estimates and comparison against treatment policy estimand under an assumed HR of 0.80 for the transition hazards of the illness-death model in the simulation with fixed-effects data-generating mechanism, fixed-effects meta-analysis, and censoring switchers*

| Scenarios | Estimators | Estimated treatment effects: HR (averaged 95% CI) | Bias (2.5%, 97.5% difference percentiles) | Coverage |
|---|---|---|---|---|
| **75% switching rate for control arm** | | | **Comparison against treatment policy estimand (True HR = 0.84)** | |
| 2:1 allocation, 75% switching rate for control arm[a] | Pure HE (100%) | 0.81 (0.69, 0.95) | -0.03 (-0.15, 0.10) | 0.92 |
| | Mixed TPE (25%) and HE (75%) | 0.82 (0.72, 0.94) | -0.02 (-0.12, 0.10) | 0.94 |
| | Mixed TPE (50%) and HE (50%) | 0.83 (0.74, 0.93) | -0.01 (-0.10, 0.09) | 0.95 |
| | Mixed TPE (75%) and HE (25%) | 0.83 (0.75, 0.93) | -0.01 (-0.09, 0.08) | 0.95 |
| | Pure TPE (100%) | 0.84 (0.76, 0.92) | -0.00 (-0.08, 0.08) | 0.95 |
| 1:1 allocation, 75% switching rate for control arm | Pure HE (100%) | 0.81 (0.70, 0.94) | -0.03 (-0.14, 0.10) | 0.93 |
| | Mixed TPE (25%) and HE (75%) | 0.83 (0.73, 0.93) | -0.01 (-0.11, 0.09) | 0.94 |
| | Mixed TPE (50%) and HE (50%) | 0.83 (0.75, 0.93) | -0.01 (-0.09, 0.09) | 0.95 |
| | Mixed TPE (75%) and HE (25%) | 0.84 (0.76, 0.92) | -0.00 (-0.08, 0.08) | 0.95 |
| | Pure TPE (100%) | 0.84 (0.77, 0.92) | 0.00 (-0.07, 0.08) | 0.95 |
| **50% switching rate for control arm** | | | **Comparison against treatment policy estimand (True HR = 0.83)** | |
| 2:1 allocation, 50% switching rate for control arm | Pure HE (100%) | 0.80 (0.71, 0.91) | -0.03 (-0.12, 0.08) | 0.92 |
| | Mixed TPE (25%) and HE (75%) | 0.81 (0.72, 0.91) | -0.02 (-0.11, 0.08) | 0.93 |
| | Mixed TPE (50%) and HE (50%) | 0.81 (0.73, 0.91) | -0.01 (-0.10, 0.08) | 0.94 |
| | Mixed TPE (75%) and HE (25%) | 0.82 (0.74, 0.91) | -0.01 (-0.09, 0.08) | 0.94 |
| | Pure TPE (100%) | 0.82 (0.75, 0.91) | -0.00 (-0.08, 0.08) | 0.95 |
| | Pure HE (100%) | 0.80 (0.72, 0.90) | -0.02 (-0.11, 0.07) | 0.92 |
| 1:1 allocation,<br>50% switching rate for control arm | Mixed TPE (25%) and HE (75%) | 0.81 (0.73, 0.90) | -0.01 (-0.10, 0.08) | 0.94 |
| | Mixed TPE (50%) and HE (50%) | 0.82 (0.74, 0.90) | -0.01 (-0.09, 0.08) | 0.94 |
| | Mixed TPE (75%) and HE (25%) | 0.82 (0.75, 0.90) | -0.00 (-0.08, 0.08) | 0.95 |
| | Pure TPE (100%) | 0.83 (0.76, 0.90) | 0.00 (-0.07, 0.08) | 0.95 |

**[a]This table shows estimated treatment effects under an assumed hazard ratio (HR) of 0.80 for the transition hazards of the illness-death model and bias and coverage in comparison to the true treatment policy estimand. Monte Carlo standard errors for all measures are very close to zero**
**Acronyms: CI: Confidence intervals; HE - Hypothetical estimator; HR - Hazard ratio; TPE - Treatment policy estimator.**

*Table 12: Averages of pooled treatment effect estimates and comparison against treatment policy estimand under an assumed HR of 1.00 for the transition hazards of the illness-death model in the simulation with fixed-effects data-generating mechanism, fixed-effects meta-analysis, and censoring switchers*

| **Scenarios** | **Estimators** | **Estimated treatment effects: HR (averaged 95% CI)** | **Bias (2.5%, 97.5% difference percentiles)** | **Coverage** |
|---|---|---|---|---|
| **75% switching rate for control arm** | | | **Comparison against treatment policy estimand (True HR = 1.00)** | |
| 2:1 allocation, 75% switching rate for control arm[a] | Pure HE (100%) | 1.01 (0.85, 1.19) | 0.01 (-0.15, 0.18) | 0.96 |
| | Mixed TPE (25%) and HE (75%) | 1.00 (0.87, 1.15) | 0.00 (-0.13, 0.14) | 0.95 |
| | Mixed TPE (50%) and HE (50%) | 1.00 (0.89, 1.13) | 0.00 (-0.11, 0.12) | 0.95 |
| | Mixed TPE (75%) and HE (25%) | 1.00 (0.90, 1.11) | -0.00 (-0.10, 0.11) | 0.95 |
| | Pure TPE (100%) | 1.00 (0.91, 1.10) | -0.00 (-0.09, 0.10) | 0.95 |
| 1:1 allocation, 75% switching rate for control arm | Pure HE (100%) | 1.02 (0.87, 1.18) | 0.02 (-0.13, 0.18) | 0.95 |
| | Mixed TPE (25%) and HE (75%) | 1.01 (0.89, 1.14) | 0.01 (-0.11, 0.14) | 0.95 |
| | Mixed TPE (50%) and HE (50%) | 1.00 (0.90, 1.12) | 0.00 (-0.10, 0.12) | 0.95 |
| | Mixed TPE (75%) and HE (25%) | 1.00 (0.91, 1.11) | 0.00 (-0.09, 0.11) | 0.95 |
| | Pure TPE (100%) | 1.00 (0.91, 1.10) | 0.00 (-0.09, 0.10) | 0.95 |
| **50% switching rate for control arm** | | | **Comparison against treatment policy estimand (True HR = 1.00)** | |
| 2:1 allocation, 50% switching rate for control arm | Pure HE (100%) | 1.00 (0.88, 1.13) | -0.00 (-0.12, 0.13) | 0.95 |
| | Mixed TPE (25%) and HE (75%) | 1.00 (0.89, 1.12) | -0.00 (-0.11, 0.12) | 0.95 |
| | Mixed TPE (50%) and HE (50%) | 1.00 (0.90, 1.11) | -0.00 (-0.11, 0.11) | 0.95 |
| | Mixed TPE (75%) and HE (25%) | 1.00 (0.90, 1.11) | -0.00 (-0.10, 0.10) | 0.95 |
| | Pure TPE (100%) | 1.00 (0.91, 1.10) | -0.00 (-0.09, 0.10) | 0.95 |
| | Pure HE (100%) | 1.00 (0.89, 1.13) | 0.00 (-0.11, 0.12) | 0.95 |
| 1:1 allocation,<br>50% switching rate for control arm | Mixed TPE (25%) and HE (75%) | 1.00 (0.90, 1.12) | 0.00 (-0.10, 0.11) | 0.95 |
| | Mixed TPE (50%) and HE (50%) | 1.00 (0.90, 1.11) | 0.00 (-0.10, 0.11) | 0.95 |
| | Mixed TPE (75%) and HE (25%) | 1.00 (0.91, 1.10) | 0.00 (-0.09, 0.10) | 0.95 |
| | Pure TPE (100%) | 1.00 (0.91, 1.10) | 0.00 (-0.09, 0.10) | 0.95 |

**[a]This table shows estimated treatment effects under an assumed hazard ratio (HR) of 1.00 for the transition hazards of the illness-death model and bias and coverage in comparison to the true treatment policy estimand. Monte Carlo standard errors for all measures are very close to zero**
**Acronyms: CI: Confidence intervals; HE - Hypothetical estimator; HR - Hazard ratio; TPE - Treatment policy estimator.**

### *3.3 Random-effects data-generating mechanism, random-effects meta-analysis*

*Table 13: Averages of pooled treatment effect estimates and comparison against treatment policy estimand under an assumed HR of 0.60 for the transition hazards of the illness-death model in the simulation with random-effects data-generating mechanism, random-effects meta-analysis, and censoring switchers*

| **Scenarios** | **Estimators** | **Estimated treatment effects: HR (averaged 95% CI)** | **Bias (2.5%, 97.5% difference percentiles)** | **Coverage** |
|---|---|---|---|---|
| **75% switching rate for control arm** | | | **Comparison against treatment policy estimand (True HR = 0.66)** | |
| 2:1 allocation, 75% switching rate for control arm[a] | Pure HE (100%) | 0.61 (0.50, 0.74) | -0.05 (-0.16, 0.08) | 0.85 |
| | Mixed TPE (25%) and HE (75%) | 0.63 (0.53, 0.75) | -0.03 (-0.14, 0.08) | 0.89 |
| | Mixed TPE (50%) and HE (50%) | 0.64 (0.55, 0.75) | -0.02 (-0.12, 0.09) | 0.91 |
| | Mixed TPE (75%) and HE (25%) | 0.65 (0.56, 0.75) | -0.01 (-0.10, 0.10) | 0.92 |
| | Pure TPE (100%) | 0.66 (0.58, 0.76) | -0.00 (-0.09, 0.10) | 0.92 |
| 1:1 allocation, 75% switching rate for control arm | Pure HE (100%) | 0.61 (0.51, 0.74) | -0.05 (-0.16, 0.08) | 0.84 |
| | Mixed TPE (25%) and HE (75%) | 0.63 (0.53, 0.75) | -0.03 (-0.13, 0.08) | 0.88 |
| | Mixed TPE (50%) and HE (50%) | 0.64 (0.55, 0.75) | -0.02 (-0.11, 0.09) | 0.91 |
| | Mixed TPE (75%) and HE (25%) | 0.65 (0.57, 0.75) | -0.01 (-0.10, 0.09) | 0.91 |
| | Pure TPE (100%) | 0.66 (0.58, 0.75) | 0.00 (-0.08, 0.10) | 0.92 |
| **50% switching rate for control arm** | | | **Comparison against treatment policy estimand (True HR = 0.64)** | |
| 2:1 allocation, 50% switching rate for control arm | Pure HE (100%) | 0.60 (0.51, 0.71) | -0.04 (-0.13, 0.07) | 0.85 |
| | Mixed TPE (25%) and HE (75%) | 0.61 (0.52, 0.72) | -0.03 (-0.12, 0.08) | 0.88 |
| | Mixed TPE (50%) and HE (50%) | 0.62 (0.53, 0.72) | -0.02 (-0.11, 0.08) | 0.90 |
| | Mixed TPE (75%) and HE (25%) | 0.63 (0.54, 0.73) | -0.01 (-0.10, 0.09) | 0.91 |
| | Pure TPE (100%) | 0.64 (0.55, 0.73) | -0.00 (-0.09, 0.10) | 0.92 |
| 1:1 allocation,<br>50% switching rate for control arm | Pure HE (100%) | 0.60 (0.51, 0.71) | -0.04 (-0.13, 0.07) | 0.85 |
| | Mixed TPE (25%) and HE (75%) | 0.61 (0.53, 0.72) | -0.03 (-0.12, 0.08) | 0.88 |
| | Mixed TPE (50%) and HE (50%) | 0.62 (0.54, 0.72) | -0.02 (-0.11, 0.09) | 0.90 |
| | Mixed TPE (75%) and HE (25%) | 0.63 (0.55, 0.73) | -0.01 (-0.09, 0.09) | 0.91 |
| | Pure TPE (100%) | 0.64 (0.56, 0.73) | 0.00 (-0.09, 0.10) | 0.92 |

**[a]This table shows estimated treatment effects under an assumed hazard ratio (HR) of 0.60 for the transition hazards of the illness-death model and bias and coverage in comparison to the true treatment policy estimand. Monte Carlo standard errors for all measures are very close to zero**
**Acronyms: CI: Confidence intervals; HE - Hypothetical estimator; HR - Hazard ratio; TPE - Treatment policy estimator.**

*Table 14: Averages of pooled treatment effect estimates and comparison against treatment policy estimand under an assumed HR of 0.80 for the transition hazards of the illness-death model in the simulation with random-effects data-generating mechanism, random-effects meta-analysis, and censoring switchers*

| Scenarios | Estimators | Estimated treatment effects: HR (averaged 95% CI) | Bias (2.5%, 97.5% difference percentiles) | Coverage |
|---|---|---|---|---|
| **75% switching rate for control arm** | | | **Comparison against treatment policy estimand (True HR = 0.84)** | |
| 2:1 allocation, 75% switching rate for control arm[a] | Pure HE (100%) | 0.81 (0.66, 0.99) | -0.03 (-0.18, 0.15) | 0.91 |
| | Mixed TPE (25%) and HE (75%) | 0.82 (0.69, 0.98) | -0.02 (-0.16, 0.14) | 0.91 |
| | Mixed TPE (50%) and HE (50%) | 0.83 (0.71, 0.97) | -0.01 (-0.14, 0.13) | 0.92 |
| | Mixed TPE (75%) and HE (25%) | 0.83 (0.72, 0.96) | -0.01 (-0.12, 0.12) | 0.92 |
| | Pure TPE (100%) | 0.84 (0.73, 0.96) | -0.00 (-0.11, 0.11) | 0.92 |
| 1:1 allocation, 75% switching rate for control arm | Pure HE (100%) | 0.81 (0.67, 0.98) | -0.03 (-0.17, 0.14) | 0.91 |
| | Mixed TPE (25%) and HE (75%) | 0.82 (0.69, 0.97) | -0.02 (-0.15, 0.13) | 0.91 |
| | Mixed TPE (50%) and HE (50%) | 0.83 (0.71, 0.96) | -0.01 (-0.13, 0.12) | 0.92 |
| | Mixed TPE (75%) and HE (25%) | 0.83 (0.73, 0.96) | -0.01 (-0.12, 0.12) | 0.91 |
| | Pure TPE (100%) | 0.84 (0.74, 0.95) | -0.00 (-0.11, 0.12) | 0.92 |
| **50% switching rate for control arm** | | | **Comparison against treatment policy estimand (True HR = 0.83)** | |
| 2:1 allocation, 50% switching rate for control arm | Pure HE (100%) | 0.80 (0.68, 0.95) | -0.02 (-0.16, 0.13) | 0.90 |
| | Mixed TPE (25%) and HE (75%) | 0.81 (0.69, 0.95) | -0.02 (-0.14, 0.12) | 0.91 |
| | Mixed TPE (50%) and HE (50%) | 0.81 (0.70, 0.95) | -0.01 (-0.13, 0.12) | 0.91 |
| | Mixed TPE (75%) and HE (25%) | 0.82 (0.71, 0.95) | -0.01 (-0.12, 0.12) | 0.92 |
| | Pure TPE (100%) | 0.82 (0.72, 0.95) | -0.00 (-0.12, 0.12) | 0.92 |
| | Pure HE (100%) | 0.80 (0.68, 0.95) | -0.02 (-0.15, 0.12) | 0.90 |

| 1:1 allocation, 50% switching rate for control arm | Mixed TPE (25%) and HE (75%) | 0.81 (0.69, 0.95) | -0.02 (-0.14, 0.12) | 0.91 |
| --- | --- | --- | --- | --- |
| | Mixed TPE (50%) and HE (50%) | 0.81 (0.70, 0.95) | -0.01 (-0.13, 0.12) | 0.92 |
| | Mixed TPE (75%) and HE (25%) | 0.82 (0.71, 0.94) | -0.01 (-0.12, 0.12) | 0.91 |
| | Pure TPE (100%) | 0.82 (0.72, 0.94) | -0.00 (-0.11, 0.12) | 0.91 |

**[a]This table shows estimated treatment effects under an assumed hazard ratio (HR) of 0.80 for the transition hazards of the illness-death model and bias and coverage in comparison to the true treatment policy estimand. Monte Carlo standard errors for all measures are very close to zero**
**Acronyms: CI: Confidence intervals; HE - Hypothetical estimator; HR - Hazard ratio; TPE - Treatment policy estimator.**

*Table 15: Averages of pooled treatment effect estimates and comparison against treatment policy estimand under an assumed HR of 1.00 for the transition hazards of the illness-death model in the simulation with random-effects data-generating mechanism, random-effects meta-analysis, and censoring switchers*

| Scenarios | Estimators | Estimated treatment effects: HR (averaged 95% CI) | Bias (2.5%, 97.5% difference percentiles) | Coverage |
|---|---|---|---|---|
| **75% switching rate for control arm** | | | **Comparison against treatment policy estimand (True HR = 1.00)** | |
| 2:1 allocation, 75% switching rate for control arm[a] | Pure HE (100%) | 1.01 (0.82, 1.24) | 0.01 (-0.18, 0.24) | 0.94 |
| | Mixed TPE (25%) and HE (75%) | 1.00 (0.84, 1.20) | 0.00 (-0.16, 0.20) | 0.93 |
| | Mixed TPE (50%) and HE (50%) | 1.00 (0.86, 1.17) | 0.00 (-0.15, 0.17) | 0.93 |
| | Mixed TPE (75%) and HE (25%) | 1.00 (0.86, 1.15) | -0.00 (-0.14, 0.15) | 0.93 |
| | Pure TPE (100%) | 1.00 (0.87, 1.14) | -0.00 (-0.13, 0.13) | 0.93 |
| 1:1 allocation, 75% switching rate for control arm | Pure HE (100%) | 1.02 (0.84, 1.24) | 0.02 (-0.16, 0.23) | 0.94 |
| | Mixed TPE (25%) and HE (75%) | 1.01 (0.85, 1.20) | 0.01 (-0.15, 0.19) | 0.93 |
| | Mixed TPE (50%) and HE (50%) | 1.00 (0.86, 1.17) | 0.00 (-0.14, 0.17) | 0.92 |
| | Mixed TPE (75%) and HE (25%) | 1.00 (0.87, 1.15) | 0.00 (-0.13, 0.15) | 0.92 |
| | Pure TPE (100%) | 1.00 (0.88, 1.14) | -0.00 (-0.12, 0.13) | 0.92 |
| **50% switching rate for control arm** | | | **Comparison against treatment policy estimand (True HR = 1.00)** | |
| 2:1 allocation, 50% switching rate for control arm | Pure HE (100%) | 1.00 (0.84, 1.19) | 0.00 (-0.16, 0.19) | 0.93 |
| | Mixed TPE (25%) and HE (75%) | 1.00 (0.85, 1.18) | -0.00 (-0.15, 0.18) | 0.92 |
| | Mixed TPE (50%) and HE (50%) | 1.00 (0.86, 1.17) | -0.00 (-0.15, 0.16) | 0.92 |
| | Mixed TPE (75%) and HE (25%) | 1.00 (0.86, 1.15) | -0.00 (-0.14, 0.15) | 0.92 |
| | Pure TPE (100%) | 1.00 (0.87, 1.15) | -0.00 (-0.13, 0.15) | 0.92 |
| | Pure HE (100%) | 1.00 (0.85, 1.18) | 0.00 (-0.16, 0.18) | 0.92 |
| 1:1 allocation,<br>50% switching rate for control arm | Mixed TPE (25%) and HE (75%) | 1.00 (0.86, 1.17) | -0.00 (-0.15, 0.17) | 0.92 |
| | Mixed TPE (50%) and HE (50%) | 1.00 (0.86, 1.16) | -0.00 (-0.14, 0.16) | 0.92 |
| | Mixed TPE (75%) and HE (25%) | 1.00 (0.87, 1.15) | -0.00 (-0.14, 0.15) | 0.92 |
| | Pure TPE (100%) | 1.00 (0.87, 1.14) | -0.00 (-0.13, 0.14) | 0.92 |

**[a]This table shows estimated treatment effects under an assumed hazard ratio (HR) of 1.00 for the transition hazards of the illness-death model and bias and coverage in comparison to the true treatment policy estimand. Monte Carlo standard errors for all measures are very close to zero**
**Acronyms: CI: Confidence intervals; HE - Hypothetical estimator; HR - Hazard ratio; TPE - Treatment policy estimator.**